\documentclass[12pt]{article}\usepackage{qSam}
\input{MyQcircuit.sty}
\begin{document}

\title{Use of a Quantum Computer to do \\
Importance and Metropolis-Hastings Sampling\\
of a Classical Bayesian Network }

\author{Robert R. Tucci\\
        P.O. Box 226\\
        Bedford,  MA   01730\\
        tucci@ar-tiste.com}

\date{ \today}

\maketitle

\vskip2cm
\section*{Abstract}
Importance sampling and
Metropolis-Hastings sampling
(of which Gibbs sampling is a special case)
are two methods
commonly used to sample multi-variate
probability distributions
(that is, Bayesian networks).
Heretofore, the sampling of
Bayesian networks has been done on
a conventional
``classical computer". In this paper,
 we  propose methods for doing
 importance sampling and
Metropolis-Hastings sampling
of a classical Bayesian network on a
quantum computer.

\section{Introduction}

Monte Carlo methods are
frequently used to
sample probability distributions.
For a
single random variable, it
is common to draw samples using the
inverse transform method\cite{inv-trans}
or
the ARM
(acceptance-rejection method)\cite{arm}.
For an n-tuple of dependent random variables
(i.e., a Bayesian network),
it is common to use
importance sampling (see Appendix
\ref{app-imp-sam}),
Gibbs sampling (see Appendix \ref{app-gibbs-sam})
and
Metropolis-Hastings sampling
(see Appendix \ref{app-met-has-sam}).
Two special cases of
importance sampling are
rejection sampling and
likelihood weighted sampling
(a.k.a. likelihood weighting).

In previous papers written by me,
I define some nets
 that describe quantum phenomena.
I call them
``quantum Bayesian nets"(QB nets).
They are a counterpart to the
conventional ``classical Bayesian
nets" (CB nets)\cite{Jordan} that describe
classical phenomena.

Heretofore, the sampling of
CB nets has been done on a conventional
 ``classical computer". In this paper,
 we advocate
 sampling a CB net with a quantum computer.

In Ref.\cite{Tuc00}, we proposed a method for
``embedding" a CB net within a
QB net. By applying this embedding technique,
we were able to obtain
in Ref.\cite{Tuc-QMR}
a method of
doing both
rejection sampling and
likelihood weighted sampling
of a CB net on a quantum
computer. In Ref.\cite{Tuc-QMR},
we  illustrated
our technique by applying it
to a special CB net used
in medical diagnosis,
the QMR (Quick Medical Reference)
CB net.

In this paper, we
generalize the results of
Ref.\cite{Tuc-QMR}
to include all kinds
of importance sampling,
(not just rejection
and likelihood weighted sampling).
We also show how to do
Gibbs sampling and
Metropolis-Hastings sampling
of a CB net with a quantum
computer.

Other workers\cite{grover, woc1,woc2,fox}
have considered
sampling of a
probability distribution
using a quantum computer.
Their methods are
very
different from ours.
Contrary to them, we
utilize a general technique,
first proposed in Ref.\cite{Tuc00},
for embedding CB nets within QB nets.
We leave to future work a deeper,
more detailed comparison between
their methods and ours.

\section{Notation and Preliminaries}

In this section, we will
define some notation that is
used throughout this paper.
For additional information about our
notation, we recommend that
the reader
consult Ref.\cite{Paulinesia}.
Ref.\cite{Paulinesia} is
a review article, written
by the author of this paper, which
uses the same notation as this paper.

We will often
use the symbol $\nb$ for the number ($\geq 1$) of qubits and
$\ns = 2^\nb$ for the number of states with $\nb$ qubits.
The quantum computing literature
often uses $n$ for $\nb$ and $N$
for $\ns$, but we will avoid this
notation. We prefer to use $n$
for the number operator, defined below.

Let $Bool =\{0, 1\}$. As usual, let $\ZZ, \RR, \CC$ represent the set
of integers (negative  and non-negative),
real numbers, and
complex numbers, respectively.
For integers $a$, $b$
 such that $a\leq b$, let
$Z_{a,b}=\{a, a+1,
\ldots b-1, b\}$.
For any positive integer $k$
and any set $S$, let
$S^k$ denote
the Cartesian product of
$k$ copies of $S$; i.e.,  the set of
all $k$-tuples
of elements of $S$.
For any set $S$, let $|S|$
be the number of elements in $S$.

We will use $\Theta(S)$
to represent the ``truth function";
$\Theta(S)$ equals 1 if statement $S$ is true
and 0 if $S$ is false.
For example, the Kronecker delta
function is defined by
$\delta^y_x=\delta(x,y) = \Theta(x=y)$.

Let
$\overline{0}=1$
and $\overline{1}=0$.
If $\veca = a_{\nb-1} \ldots a_2 a_1 a_0$,
where $a_\mu\in Bool$, then
$dec(\veca) = \sum^{\nb-1}_{\mu=0} 2^\mu a_\mu=a$.
Conversely, $\veca=bin(a)$.

We define the single-qubit states $\ket{0}$ and $\ket{1}$ by

\beq
\ket{0} =
\left[
\begin{array}{c}
1 \\ 0
\end{array}
\right]
\;\;,\;\;
\ket{1} =
\left[
\begin{array}{c}
0 \\ 1
\end{array}
\right]
\;.
\eeq
If $\veca \in Bool^{\nb}$, we define the
$\nb$-qubit state $\ket{\veca}$ as the following tensor product

\beq
\ket{\veca} = \ket{a_{\nb -1}}
\otimes \ldots \ket{a_1} \otimes \ket{a_0}
\;.
\eeq
For example,

\beq
\ket{01} =
\left[
\begin{array}{c}
1 \\ 0
\end{array}
\right]
\otimes
\left[
\begin{array}{c}
0 \\ 1
\end{array}
\right]
=
\left[
\begin{array}{c}
0 \\ 1 \\ 0 \\0
\end{array}
\right]
\;.
\eeq

When we write a matrix, and
leave some of its entries
blank, those blank entries
should be interpreted as zeros.

$I_k$ and $0_k$ will represent the
$k\times k$ unit and zero matrices, respectively.
For any matrix $A\in\CC^{p\times q}$,
$A^*$ will stand for its complex
conjugate, $A^T$ for its transpose, and
$A^\dagger$ for its Hermitian conjugate.

For any matrix $A$ and positive integer $k$,
let

\beq
A^{\otimes k} =
\underbrace{A\otimes \cdots
\otimes A \otimes A}_{k \mbox{\tiny\;\;copies of } A}
\;,
\eeq

\beq
A^{\oplus k} =
\underbrace{A\oplus \cdots
\oplus A \oplus A}_{k \mbox{\tiny\;\;copies of } A}
\;.
\eeq

Suppose $\beta\in Z_{0,\nb-1}$ and
$M$ is any $2\times 2$ matrix. We define
$M(\beta)$ by

\beq
M(\beta) =
I_2 \otimes
\cdots \otimes
I_2 \otimes
M \otimes
I_2 \otimes
\cdots \otimes
I_2
\;,
\label{eq-m-beta-def}
\eeq
where the matrix $M$ on the right
hand side is located
at qubit position $\beta$ in the tensor product
of $\nb$ $2\times 2$ matrices.
The numbers that label qubit positions in the
tensor product increase from
right to left ($\leftarrow$),
and the rightmost qubit is taken
to be at position 0.

The Pauli matrices
are

\beq
\sigx=
\left(
\begin{array}{cc}
0&1\\
1&0
\end{array}
\right)
\;,
\;\;
\sigy=
\left(
\begin{array}{cc}
0&-i\\
i&0
\end{array}
\right)
\;,
\;\;
\sigz=
\left(
\begin{array}{cc}
1&0\\
0&-1
\end{array}
\right)
\;.
\eeq
Let $\vec{\sigma} = (\sigx, \sigy , \sigz)$.
For any $\veca\in\RR^3$,
let $\sigma_{\veca}=\vec{\sigma}\cdot\veca$.

The one-qubit Hadamard
matrix $H$ is defined as:

\beq
H= \frac{1}{\sqrt{2}}
\left[
\begin{array}{cc}
1&1\\
1&-1
\end{array}
\right]
\;.
\eeq
The $\nb$-qubit Hadamard matrix is defined
as $H^{\otimes \nb}$.

The number operator $n$ for
a single qubit is defined by

\beq
n =
\left[
\begin{array}{cc}
0 & 0 \\
0 & 1
\end{array}
\right]
=
\frac{ 1 - \sigz}{2}
\;.
\eeq
Note that

\beq
n \ket{0} = 0\ket{0} = 0
\;\;,\;\;
n \ket{1} = 1\ket{1}
\;.
\eeq
We will often use $\nbar$ as shorthand for

\beq
\nbar =
1-n =
\left[
\begin{array}{cc}
1 & 0 \\
0 & 0
\end{array}
\right]
=
\frac{ 1 + \sigz}{2}
\;.
\eeq
Define $P_0$ and $P_1$ by

\beq
P_0 = \nbar =
\left[
\begin{array}{cc}
1 & 0 \\
0 & 0
\end{array}
\right]=
\ket{0}\bra{0}
\;\;,\;\;
P_1 = n =
\left[
\begin{array}{cc}
0 & 0 \\
0 & 1
\end{array}
\right]
=
\ket{1}\bra{1}
\;.
\eeq
$P_0$ and $P_1$ are orthogonal projection
operators and they add to one:

\beq
P_a P_b = \delta(a, b) P_b
\;\;\;\;\; {\rm for} \;\; a,b\in Bool
\;,
\eeq

\beq
P_0 +  P_1 = I_2
\;.
\eeq

For $\veca \in Bool^\nb$, let

\beq
P_{\veca} = P_{a_{\nb-1}} \otimes \cdots
\otimes P_{a_2} \otimes P_{a_1} \otimes P_{a_0}
\;.
\eeq
For example,
with 2 qubits we have

\beq
P_{00} = P_0 \otimes P_0 = diag(1, 0, 0, 0)
\;,
\eeq

\beq
P_{01} = P_0 \otimes P_1 = diag(0, 1, 0, 0)
\;,
\eeq

\beq
P_{10} = P_1 \otimes P_0 = diag(0, 0, 1, 0)
\;,
\eeq

\beq
P_{11} = P_1 \otimes P_1 = diag(0, 0, 0, 1)
\;.
\eeq
Note that

\beq
P_\veca P_\vecb = \delta(\veca, \vecb) P_\vecb
\;\;\;\;\; {\rm for} \;\; \veca,\vecb\in Bool^\nb
\;,
\eeq

\beq
\sum_{\veca\in Bool^\nb }
P_{\veca} =
I_2 \otimes I_2 \otimes \cdots \otimes I_2 = I_{2^\nb}
\;.
\eeq

Next we explain our circuit diagram notation.
We label single qubits (or qubit
positions) by a Greek letter or by an
integer. When we use integers,
the topmost qubit wire is 0, the next one
down is 1, then
2, etc.
{\it Note that in our circuit diagrams,
time flows from the right to the left
of the diagram.} Careful:
Many workers in Quantum
Computing draw their diagrams
so that time flows from
left to right. We eschew their
convention because
it forces one to reverse
the order of the operators
every time one wishes to convert
between a circuit
diagram
and its algebraic equivalent
in Dirac notation.

Suppose $U\in U(2)$.
If $\tau$ and $\kappa$ are two different
qubit positions, gate
$U(\tau)^{n(\kappa)}$ (or $U(\tau)^{\nbar(\kappa)}$)
is called a {\bf controlled $U$}
with target $\tau$ and control $\kappa$.
When $U=\sigx$, this reduces to a
{\bf CNOT (controlled NOT)}.
 If
 $\tau$,$\kappa_1$ and $\kappa_0$
are 3 different qubit positions,
$\sigx(\tau)^{n(\kappa_1)n(\kappa_0)}$
is called a {\bf Toffoli gate} with target
$\tau$ and controls $\kappa_1, \kappa_0$.
Suppose $N_K\geq 2$ is an integer and
$\vec{b}\in Bool^{N_K}$.
Suppose
$\tau, \kappa_{N_K-1},\kappa_{N_K-2},
\ldots,\kappa_1,\kappa_0$ are distinct qubits
and $\vec{\kappa}=
(\kappa_{N_K-1},\kappa_{N_K-2},
\ldots,\kappa_1,\kappa_0)$.
Gate
$U(\tau)^{P_{\vec{b}}
(\vec{\kappa})}$
is called a {\bf multiply
controlled $U$}
with target $\tau$ and $N_K$
controls
$\vec{\kappa}$.
When $U=\sigx$, this reduces to an
{\bf MCNOT (multiply controlled
NOT)}.

For any set $\Omega$ and any
function $f:\Omega\rarrow \CC$,
we will use
$f(x)/(\sum_{x\in\Omega} num)$,
where ``num" stands for numerator, to mean
$f(x)/(\sum_{x\in\Omega} f(x))$.
This notation is convenient when
$f(x)$ is a long expression
that we do not wish to write twice.

Consider an n-tuple $\vecf=(f_1,f_2, \ldots, f_n)$,
and a set $A\subset Z_{1,n}$.
By $(\vecf)_A$ we will mean $(f_i)_{i\in A}$
;
that is,
the $|A|$-tuple
that one creates from $\vecf$,
by keeping only the components
listed in $A$.

Symbols which represent random variables
will be underlined.
The set of values (or states) that a random
variable $\rvx$
can assume will be denoted by $val(\rvx)$ (
or $S_\rvx$). Samples of $\rvx$ will be denoted
by $\sam{x}{k}$ for $k\in Z_{1, N_{sam}}$.

Next, consider a CB net with nodes
$\rvx_1, \rvx_2, \ldots,\rvx_{N_{nds}}$.

We will use $pa(i)$ ($ch(i)$, respectively)
to denote the
set of all $j\in Z_{1,N_{nds}}$
such that $\rvx_j$
is a parent (child, respectively) of $\rvx_i$.
Suppose $\gamma = pa, ch$.
Let $\gamma(\rvx_i)=\{\rvx_j:j\in \gamma(i)\}$.
Let $\gamma(S)=\cup_{i\in S} \gamma(i)$.

\begin{figure}[h]
    \begin{center}
    \epsfig{file=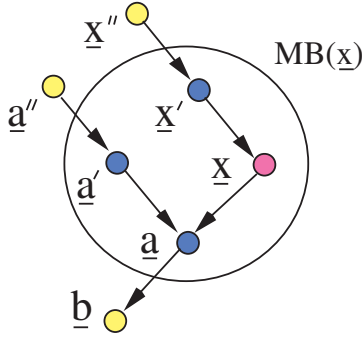, height=1.8in}
    \caption{The Markov blanket of node $\rvx$
    is the set of all nodes inside the large circle, excluding $\rvx$.}
    \label{fig-markov-blanket}
    \end{center}
\end{figure}

The
{\bf Markov blanket} of $\rvx_i$ is defined by

\beq
MB(i) = pa(i) \cup ch(i) \cup pa(ch(i))
\;.
\eeq
Let $\{i\}^c=Z_{1,N_{nds}}-\{i\}$.
One an prove that

\beq
P(x_i| (x)_\noti)=
P(x_i| (x)_{MB(i)})
\;.
\label{eq-mb}
\eeq
We won't prove Eq.(\ref{eq-mb}) here,
but next we will give an example
to make it plausible.
For the CB net shown in Fig.\ref{fig-markov-blanket},
one has

\beqa
P(x|x', x'',a,a',a'',  b)&=&
\frac{P(x|x')P(x'|x'')P(x'')P(a|a',x)P(a'|a'')P(a'')P(b|a)}
{\sum_x num}\nonumber\\
&=&
\frac{P(x|x')P(a|a',x)}
{\sum_x num}\nonumber\\
&=& P(x|x',a,a')
\;.
\eeqa

\section{Multiplexors}\label{sec-muxors}

In this section, we discuss
 some multi-qubit
transformations called
multiplexors.

Suppose that $U$ is an $N\times N$
unitary matrix, where $N$ is an even number.
The {\bf Cosine-Sine Decomposition (CSD) Theorem}\cite{Golub}
states\footnote{Actually, this
is only a special case of the CSD
Theorem---the case which is
most relevant to quantum computing.
The general version of the
CSD Theorem does not restrict
the dimension of $U$ to be even,
or even restrict the blocks
into which $U$ is
partitioned to be of
equal size.}
that one can always
express $U$ in the form

\begin{subequations}
\label{eq-csd-long}
\beq
U =
\left[
\begin{array}{cc}
L_0&0\\
0&L_1
\end{array}
\right]
D
\left[
\begin{array}{cc}
R_0&0\\
0&R_1
\end{array}
\right]
\;,
\eeq
where the left and right matrices
$L_0,L_1,R_0,R_1$ are
$\frac{N}{2}\times \frac{N}{2}$
unitary matrices, and

\beq
D=
\left[
\begin{array}{cc}
D_{00}&D_{01}\\
D_{10}&D_{11}
\end{array}
\right]
\;,
\eeq

\beq
D_{00}=D_{11}=diag(C_1,C_2, \dots,C_{\frac{N}{2}})
\;,
\eeq

\beq
D_{01}=diag(S_1,S_2, \dots,S_{\frac{N}{2}})
\;,\;\;D_{10}=-D_{01}
\;.
\eeq
\end{subequations}
For all $i\in Z_{1,\frac{N}{2}}$,
\;\;$C_i=\cos\theta_i$
and
$S_i=\sin\theta_i$
for some angle $\theta_i$.
Eqs.(\ref{eq-csd-long}) can be expressed more
succinctly as

\beq
U=(L_0\oplus L_1)
e^{i\sigy\otimes \Theta}
(R_0\oplus R_1)
\;,
\eeq
 where
$\Theta=diag(\theta_1,
\theta_2, \dots, \theta_{\frac{N}{2}})$.

We will henceforth refer to Ref.\cite{Tuc99}
as Tuc99. Tuc99 was the first
paper to use the CSD
to compile unitary matrices.
By ``compiling a unitary matrix",
we mean decomposing it into a SEO (Sequence of
Elementary Operators),
elementary operators such as
single-qubit rotations and CNOTs.

Note that for some $\phi_\vecb\in \RR$
and $N=2^\nb$, matrix $D$
of
Eq.(\ref{eq-csd-long})
can be expressed as

\begin{subequations}
\label{eq-d-nb-bits}
\begin{eqnarray}
D&=&
\exp\left(i\sigy\otimes
\sum_{\vecb\in Bool^{\nb-1}}
\phi_\vecb P_\vecb\right)
\label{eq-d-nb-bits-a}
\\&=&
\sum_{\vecb\in Bool^{\nb-1}}
e^{i\phi_\vecb \sigy}\otimes P_\vecb
\label{eq-d-nb-bits-b}
\\&=&
\prod_{\vecb\in Bool^{\nb-1}}
e^{i\phi_\vecb \sigy\otimes P_\vecb}
\;.
\label{eq-d-nb-bits-c}
\end{eqnarray}
\end{subequations}
To prove that Eqs.(\ref{eq-d-nb-bits-a}),
(\ref{eq-d-nb-bits-b}), and (\ref{eq-d-nb-bits-c})
are equivalent, just
apply $\ket{\vecb}_{\nb-2,\ldots,1,0}$ with
$\vecb\in Bool^{\nb-1}$
to the right hand side of each line, and
use the fact that
$P_{\vecb'}\ket{\vecb} =
\delta_\vecb^{\vecb'}\ket{\vecb}$.
(Note that we can  ``pull the
$\vecb$ sum" out of the argument of the
exponential only if  we also pull out the $\otimes P_\vecb$.)

In Tuc99, I refer to matrices of the form
of the $D$ matrix   of
Eq.(\ref{eq-csd-long}) simply as ``D-matrices".
In my papers that followed Tuc99,
I've begun calling such matrices
``multiplexors".\footnote{``multiplexor"
means ``multi-fold" in Latin.
A special type of electronic
device is also called a multiplexor
or multiplexer.}
When I want to be more
precise, I call the $D$ matrix   of
Eq.(\ref{eq-csd-long}), an
$R_y(2)$-multiplexor with target qubit $\nb-1$
and control qubits
$\nb-2,\dots,2,1,0$. The $R_y(2)$
term refers to the fact that
the set of operations acting
on the target qubit are $2\times 2$
qubit rotations $R_y(\phi)=e^{i\phi\sigy}$
for some $\phi\in\RR$. More
generally, one can speak
of $U(N)$-multiplexors. Henceforth in this paper,
I'll continue using this multiplexor
nomenclature, even though it's
not used in Tuc99.

\begin{figure}[h]
    \begin{center}
    \epsfig{file=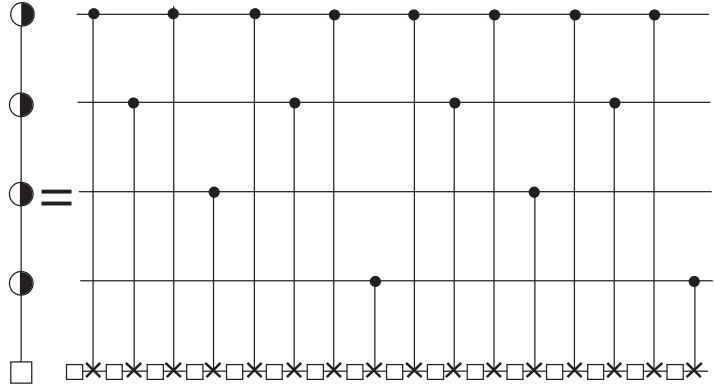, height=2.0in}
    \caption{
    A possible decomposition
    of an $R_y(2)$-multiplexor
    with 4 controls.
    }
    \label{fig-4controls}
    \end{center}
\end{figure}

Tuc99 gives identities for
decomposing an arbitrary
$R_y(2)$-multiplexor with $\nb-1$ controls
into a SEO with $2^{\nb-1}$ CNOTs.
Fig.\ref{fig-4controls} shows an example
of the SEO decomposition
found in Tuc99 for an $R_y(2)$-multiplexor. In Fig.\ref{fig-4controls},
0,1,2,3 are the control qubits,
and 4 is the target qubit. The empty
square vertices represent $R_y(2)$
gates. The symbol to the left of the equal sign,
the one with
the ``half-moon" vertices, was invented
by the authors of Ref.\cite{chain}
 to represent
a multiplexor.

$U(N)$ multiplexors for any $N\geq 2$
satisfy certain simple properties that
we shall discuss next.

Let ``con" stand for control and ``tar" for
target.

Suppose $\vecb\in Bool^{N_B,con}$
and $\{U_\vecb\}_{\forall \vecb}$
is a family of $2^{N_{B,tar}}\times
2^{N_{B,tar}}$ unitary matrices.
Define
\beq
E_{\vecb}[U_\vecb] =
\sum_\vecb U_\vecb\otimes P_\vecb
\;.
\eeq

Suppose $\vecb\in Bool^{N_B,con1}$,
$\vecb'\in Bool^{N_B,con2}$,
and $\{U_{\vecb,\vecb'}\}_{\forall \vecb,\vecb'}$
is a family of $2^{N_{B,tar}}\times
2^{N_{B,tar}}$ unitary matrices.
Define

\beq
E_{\vecb|\vecb'}[U_{\vecb,\vecb'}]=
\sum_\vecb e^{U_{\vecb,\vecb'}}\otimes P_\vecb
\;.
\eeq

\begin{claim}
If $\vecb\in Bool^{N_B,con}$
and $\{U_\vecb\}_{\forall \vecb}$,
$\{V_\vecb\}_{\forall \vecb}$
are two families of $2^{N_{B,tar}}\times
2^{N_{B,tar}}$ unitary matrices, then

\beq
E_\vecb[U_\vecb]\;\;
E_{\vecb}[V_\vecb]
=
E_\vecb[U_\vecb V_\vecb]
\;.
\eeq
The last
equation can be represented
in circuit notation. For
example, when $N_{B,con}=2$
and  $N_{B,tar}=2$, one writes

\beq
\begin{array}{c}
\Qcircuit @C=1em @R=1em @!R{
&\muxorgate
&\muxorgate
&\qw
\\
&\muxorgate\qwx[-1]
&\muxorgate\qwx[-1]
&\qw
\\
&\multigate{1}{\;}\qwx[-1]
&\multigate{1}{\;}\qwx[-1]
&\qw
\\
&\ghost{\;}
&\ghost{\;}
&\qw
}
\end{array}
=
\begin{array}{c}
\Qcircuit @C=1em @R=1em @!R{
&\muxorgate
&\qw
\\
&\muxorgate\qwx[-1]
&\qw
\\
&\multigate{1}{\;}\qwx[-1]
&\qw
\\
&\ghost{\;}
&\qw
}
\end{array}
\;.
\eeq
\end{claim}
\proof
Obvious.
\qed

\begin{claim}
If $\vecb\in Bool^{N_B,con1}$,
$\vecb'\in Bool^{N_B,con2}$, and
$\{U_{\vecb,\vecb'}\}_{\forall \vecb,\vecb'}$,
is a family of $2^{N_{B,tar}}\times
2^{N_{B,tar}}$ unitary matrices, then

\beq
E_{\vecb'}\left[E_{\vecb|\vecb'}[U_{\vecb,\vecb'}]\right]
=
E_{\vecb,\vecb'}[U_{\vecb,\vecb'}]
\;.
\eeq
The last
equation can be represented
in circuit notation. For
example, when $N_{B,con1}=1$, $N_{B,con2}=2$,
and  $N_{B,tar}=2$, one writes

\beq
\begin{array}{c}
\Qcircuit @C=1em @R=.5em @!R{
&\qw
&\muxorgate
&\qw
&\qw
\\
&\qw
&\muxorgate\qwx[-1]\qwx[1]
&\qw
&\qw
\\
&
&
\\
&\qw
&\muxorgate
&\qw
&\qw
\\
&\qw
&\multigate{1}{\;}\qwx[-1]
&\qw
&\qw
\\
&\qw
&\ghost{\;}
&\qw
&\qw
\gategroup{4}{3}{6}{3}{1.8em}{-}
}
\end{array}
=
\begin{array}{c}
\Qcircuit @C=1em @R=.5em @!R{
&\muxorgate
&\qw
\\
&\muxorgate\qwx[-1]
&\qw
\\
&\muxorgate\qwx[-1]
&\qw
\\
&\multigate{1}{\;}\qwx[-1]
&\qw
\\
&\ghost{\;}
&\qw
}
\end{array}
\;.
\eeq
\end{claim}
\proof
Obvious.
\qed

We end this section by proving a ``chain rule"
for $R_y(2)$ multiplexors (similar to
the Chapman-Kolgomorov chain rule for
conditional probabilities).
A result similar to the next
claim is given in Ref.\cite{chain}.

Below, when an index is replaced by a dot,
we
mean that the index is summed over
all its possible values. For example,
$q_{\cdot \cdot k} = \sum_{i,j} q_{ijk}$.

\begin{claim}\label{cl-muxor-chain}
Suppose
$\vecb=(b_{\nb-1}, \ldots, b_1,b_0)\in Bool^\nb$, $q_\vecb \geq 0$, $\sum_\vecb q_\vecb = 1$
Assume $\nb=3$ for definiteness. One has

\begin{subequations}
\label{eq-muxor-chain}
\beq
\sum_{\vecb\in Bool^3} \sqrt{q_\vecb} \;\ket{\vecb}=
U\ket{0}^{\otimes 3}
\;
\eeq
if

\beq
U =
\left[e^{i\sum_{b_1, b_0}\theta_{b_1 b_0}\sigy\otimes P_{b_1 b_0}}\right]
\left[I_2\otimes e^{i\sum_{b_0}\theta_{b_0}\sigy\otimes P_{b_0}}\right]
\left[I_2^{\otimes 2}\otimes e^{i\theta\sigy}\right]
\;,
\eeq
\end{subequations}
and the angles $\theta_{b_1 b_0}$,
$\theta_{b_0}$ and $\theta$
are defined by

\begin{subequations}
\beq
(C_{b_1 b_0}, S_{b_1 b_0}) =
\frac{1}{\sqrt{q_{\cdot b_1 b_0}}}
(\sqrt{q_{0 b_1 b_0}},
\sqrt{q_{1 b_1 b_0}})
\;,
\eeq

\beq
(C_{b_0}, S_{b_0}) =
\frac{1}{\sqrt{q_{\cdot \cdot b_0}}}
(\sqrt{q_{\cdot 0 b_0}},
\sqrt{q_{\cdot 1 b_0}})
\;,
\eeq

\beq
(C, S) =
(\sqrt{q_{\cdot \cdot 0}},
\sqrt{q_{\cdot \cdot 1}})
\;,
\eeq
\end{subequations}
where $C_\gamma = \cos(\theta_\gamma)$
and $S_\gamma = \sin(\theta_\gamma)$
for any symbol $\gamma$ (including no symbol).
Eq.(\ref{eq-muxor-chain}) can be represented as
a circuit diagram, as follows:

\beq
\sum_{\vecb\in Bool^3} \sqrt{q_\vecb}\;\ket{\vecb}=
\begin{array}{c}
\Qcircuit @C=1em @R=1em @!R{
&\muxorgate
&\muxorgate
&\emptygate
&\rstick{\ket{0}}\qw
\\
&\muxorgate\qwx[-1]
&\emptygate\qwx[-1]
&\qw
&\rstick{\ket{0}}\qw
\\
&\emptygate\qwx[-1]
&\qw
&\qw
&\rstick{\ket{0}}\qw
}
\end{array}
\;.
\eeq

\end{claim}
\proof

$\bra{b}P_{b'} = \delta_b^{b'} \bra{b}$ for $b,b'\in Bool$.
Thus

\beqa
\bra{b_2,b_1,b_0}U\ket{0}^{\otimes{3}} &=&
\bra{b_2} e^{i\sigy\theta_{b_1 b_0}}\ket{0}
\bra{b_1} e^{i\sigy\theta_{b_0}}\ket{0}
\bra{b_0} e^{i\sigy\theta}\ket{0}\\
&=&
[C_{b_1 b_0}^{\overline{b_2}}S_{b_1 b_0}^{b_2}]
[C_{b_0}^{\overline{b_1}}S_{b_0}^{b_1}]
[C^{\overline{b_0}}S^{b_0}]
\;.
\eeqa
Let

\begin{subequations}
\label{eq-pb-cond}
\beq
P(b_2|b_1,b_0) =
C_{b_1 b_0}^{2\overline{b_2}}S_{b_1 b_0}^{2b_2}
\;,
\eeq

\beq
P(b_1|b_0) =
C_{b_0}^{2\overline{b_1}}S_{b_0}^{2b_1}
\;,
\eeq

\beq
P(b_0) =
C^{2\overline{b_0}}S^{2b_0}
\;.
\eeq
\end{subequations}
Then

\beqa
q_\vecb &=& \left| \bra{b_2,b_1,b_0} U \ket{0}^{\otimes 3} \right|^2\\
&=& P(b_2|b_1,b_0)P(b_1|b_0)P(b_0)\\
&=& P(\vecb)
\;.
\eeqa
Eqs.(\ref{eq-pb-cond}) are satisfied if

\begin{subequations}
\beq
\frac{\sqrt{q_{b_2 b_1 b_0}}}
{\sqrt{q_{\cdot b_1 b_0}}}
 =
C_{b_1 b_0}^{\overline{b_2}}S_{b_1 b_0}^{b_2}
\;,
\eeq

\beq
\frac{\sqrt{q_{\cdot b_1 b_0}}}
{\sqrt{q_{\cdot \cdot b_0}}}
 =
C_{b_0}^{\overline{b_1}}S_{b_0}^{b_1}
\;,
\eeq

\beq
\sqrt{q_{\cdot \cdot b_0}}
=
C^{\overline{b_0}}S^{b_0}
\;.
\eeq
\end{subequations}
\qed

The above claim can be easily generalized to arbitrary $\nb>0$.

Note that $\sum_{\vecb} \sqrt{q_\vecb}\;\ket{\vecb}$,
when expressed in matrix notation, is the
column vector with entries $\sqrt{q_\vecb}$.
For example, for $\nb=3$,
it equals
$[\sqrt{q_{000}},\sqrt{q_{001}},\sqrt{q_{010}},\ldots,\sqrt{q_{111}}]^T$.

\section{Q-Embeddings}\label{sec-q-emb}
In this section, we will
review and extend
the section of Ref.\cite{Tuc00}
entitled ``Q-Embeddings".

A {\bf probability matrix} $P(y|x)$ is
a rectangular (not necessarily square)
matrix with row index $y\in S_\rvy$ and column
index $x\in S_\rvx$ such that $P(y|x)\geq 0$ for all $x,y$,
and $\sum_y P(y|x) = 1$ for all $x$.
A probability matrix is assigned to each node of a
CB net.

A unitary matrix $A(y, \tilde{x} | x, \tilde{y})$
(with rows labelled
by $y, \tilde{x}$ and columns by $x, \tilde{y}$)
is a {\bf q-embedding of probability
matrix} $P(y|x)$ if

\beq
\sum_{\tilde{x}} | A(y, \tilde{x} | x,
\tilde{y}=0) |^2 = P(y|x)
\;
\label{eq:q-embed-mat-def}\eeq
for all possible values of $y$ and $x$. (the ``q" in
``q-embedding" stands for ``quantum").
When considering a q-embedding
$A(y, \tilde{x} | x,\tilde{y})$
of a probability matrix, we
will refer to
$y$ as the {\bf focus index},
$\tilde{y}$ as the {\bf focus-image index}
or {\bf source index},
$x$ as the {\bf parent index},
and
$\tilde{x}$ as the {\bf parent-image index}
or  {\bf sink index}.
We will also refer to $\tilde{x}$ and $\tilde{y}$
collectively as
{\bf ancilla indices}.

Given a QB net $\qbnet$,
let

\beq
P[ (x)_L] =
\left| \sum_{(x)_{\Gamma_Q-L}}
A(x)
\right|^2
\;.
\label{eq:q-embed-net-predef}\eeq
On the right hand side of
Eq.(\ref{eq:q-embed-net-predef}),
$A(x)$ is the amplitude of story $x$,
$\Gamma_Q$  is the set of indices
of all the
nodes of $\qbnet$,
and
$L$ is the set of indices of
 all leaf (a.k.a. external) nodes
of $\qbnet$.
We say $\qbnet$ is a {\bf q-embedding
of CB net} $\cbnet$ if
$P[ (x)_L]$ defined by
Eq.(\ref{eq:q-embed-net-predef}) satisfies

\beq
P[ (x)_{\Gamma_C}]
=
\sum_{L_1} P[ (x)_L]
\;,
\label{eq:q-embed-net-def}
\eeq
where $L_1\subset L$,
and $\Gamma_C$ is the set of
indices of all nodes of $\cbnet$.
Thus, the probability distribution associated
with all nodes of $\cbnet$ can be obtained
from the probability distribution associated
with the external nodes of $\qbnet$.
Ref.\cite{Tuc00} gives
two examples  of q-embeddings of CB nets:
the two-body scattering net and the Asia net.
Ref.\cite{Tuc-QMR} gives the example
of the Quick Medical Reference net.
More examples will be given later in
this paper.

\subsection{Q-Embeddings of Probability Matrices}
\label{sec-q-emb-mats}

In Ref.\cite{Tuc00}, we
showed that any probability matrix has
a q-embedding. Our proof was
constructive and relied on the Gram-Schmidt
method. In this section, we will
give a new proof, again constructive,
that relies on multiplexors. The
q-embeddings constructed in
this section, compared with those
of Ref.\cite{Tuc00}, have the
advantage that they are already compiled,
whereas those obtained in
 Ref.\cite{Tuc00} in general
require  a computer program, ``a compiler",
 to compile them numerically.

Eq.(\ref{eq:q-embed-mat-def}) is satisfied by

\beq
A(y,\tilde{x}|\tilde{y}=0,x)=
\sqrt{P(y|x)}\;\;\delta_x^{\tilde{x}}
\;.
\label{eq-amp-tensor-form}
\eeq
When speaking of a parent
index $\rvx$  and a focus index $\rvy$, we will denote
number of parent bits by $N_{B,par}= \log_2(N_\rvx)$ ,
and number of focus bits by $N_{B,foc}= \log_2(N_\rvy)$,
Also,
$N_{S,par} = 2^{N_{B,par}}=N_\rvx$
, and
$N_{S,foc} = 2^{N_{B,foc}}=N_\rvy$.
Eq.(\ref{eq-amp-tensor-form})
can be expressed in matrix form as follows:

\beq
[A(y,\tilde{x}|\tilde{y}=0,x)]=
\begin{array}{c|c}
&(\tilde{y}=0,x)\rarrow\\
\hline
(y,\tilde{x})&D^{0,0}\\
\downarrow&D^{1,0}\\
&\cdots \\
&D^{N_\rvy-1,0}
\end{array}
\;,
\label{eq-amp-mat-form-one-col}
\eeq
where,
for all $y\in val(\rvy)$,
 $D^{y,0}\in \RR^{N_\rvx\times N_\rvx}$
 are diagonal matrices
 with entries

 \beq
 (D^{y,0})_{x,\tilde{x}}=\sqrt{P(y|x)}\delta_x^{\tilde{x}}
 \;.
 \eeq
 By adding
 more columns to the matrix of
 Eq.(\ref{eq-amp-mat-form-one-col}),
 one can
extended it to the following
square matrix:

\beqa
[A(y,\tilde{x}|\tilde{y},x)]&=&
\begin{array}{c|cccc}
&&(\tilde{y},x)&\rarrow&\\
\hline
(y,\tilde{x})&D^{0,0} & D^{0,1} & \cdots & D^{0,N_\rvy-1}\\
\downarrow&D^{1,0} & D^{1,1} & \cdots & D^{1,N_\rvy-1}\\
&\cdots & \cdots & \cdots & \cdots\\
&D^{N_\rvy-1,0} & D^{N_\rvy-1,1} & \cdots & D^{N_\rvy-1,N_\rvy-1}
\end{array}
\\
&=&\sum_{\vecb\in Bool^{\log_2 N_\rvx}}
U_\vecb\otimes P_\vecb
\;.
\eeqa
For all $y_1,y_2\in val(\rvy)$,
 $D^{y_1,y_2}\in \RR^{N_\rvx\times N_\rvx}$
 are diagonal
matrices. These diagonal matrices are chosen so
that
$U_\vecb\in \RR^{N_\rvy\times N_\rvy}$
are unitary matrices
such that
the first column of
$U_\vecb$ is given by
$(U_\vecb)_{y,0} = \sqrt{P(y|\rvx=dec(\vecb))}$.
The other columns of the
$U_\vecb$'s
can be chosen at will provided that
they make the $U_\vecb$'s
unitary.
According to Claim \ref{cl-muxor-chain}, we
can choose each $U_\vecb$ to be
a chain of
multiplexors,
in which case
$[A(y,\tilde{x}|\tilde{y},x)]$ is
a multiplexor of a
chain of multiplexors.
For example, if $\log_2 N_\rvy=3$
and
$\log_2 N_\rvx=2$, then

\beq
U_\vecb=
\begin{array}{c}
\Qcircuit @C=1em @R=.5em @!R{
&\muxorgate
&\muxorgate
&\emptygate
&\qw
\\
&\muxorgate\qwx[-1]
&\emptygate\qwx[-1]
&\qw
&\qw
\\
&\emptygate\qwx[-1]
&\qw
&\qw
&\qw
}
\end{array}
\;,
\eeq
and

\beq
[A(y,\tilde{x}|\tilde{y},x)]=
\begin{array}{c}
\Qcircuit @C=1em @R=.5em @!R{
&\qw
&\qw
&\muxorgate
&\qw
&\qw
&\qw
\\
&\qw
&\qw
&\muxorgate\qwx[-1]\qwx[1]
&\qw
&\qw
&\qw
\\
&
&
&
&
&
&
\\
&\qw
&\muxorgate
&\muxorgate
&\emptygate
&\qw
&\qw
\\
&\qw
&\muxorgate\qwx[-1]
&\emptygate\qwx[-1]
&\qw
&\qw
&\qw
\\
&\qw
&\emptygate\qwx[-1]
&\qw
&\qw
&\qw
&\qw
\gategroup{4}{3}{6}{5}{2.2em}{-}
}
\end{array}
=
\begin{array}{c}
\Qcircuit @C=1em @R=.5em @!R{
&\muxorgate
&\muxorgate
&\muxorgate
&\qw
\\
&\muxorgate\qwx[-1]
&\muxorgate\qwx[-1]
&\muxorgate\qwx[-1]
&\qw
\\
&\muxorgate\qwx[-1]
&\muxorgate\qwx[-1]
&\emptygate\qwx[-1]
&\qw
\\
&\muxorgate\qwx[-1]
&\emptygate\qwx[-1]
&\qw
&\qw
\\
&\emptygate\qwx[-1]
&\qw
&\qw
&\qw
}
\end{array}
\;.
\eeq
As mentioned in Section \ref{sec-muxors}
on multiplexors,
Ref.\cite{Tuc99} shows how
to decompose a multiplexor into
a SEO. Thus, this particular
q-embedding of $P(y|x)$
comes already compiled (decomposed
 into a SEO).

\subsection{Q-Embeddings of CB nets}

Ref.\cite{Tuc00} describes
a method by which, given
 any CB net $\cbnet$, one can construct a QB net
 $\qbnet$ which is a q-embedding of $\cbnet$.
 In this section, we will review this method.
 The method will be used in later sections
 to construct QB nets for sampling.

In the previous section, we showed how to
 construct a q-embedding for
any probability matrix. Now remember that each
node of $\cbnet$ has a probability
matrix assigned to it.
The main step in constructing
a q-embedding of $\cbnet$ is to
replace each node matrix of $\cbnet$ with
a q-embedding of it.

Before describing our construction method,
we need some definitions.
We say a node $\rvm$
 is a {\bf marginalizer node} if it
has a single input arrow and a single
output arrow.
Furthermore, the parent node of $\rvm$,
call it $\rvx$,
has states $x=(x_1, x_2, \ldots, x_n)$,
where  $x_i\in S_{\rvx_i}$ for each $i\in Z_{1,n}$.
Furthermore, for some particular integer
$i_0\in Z_{1,n}$,
the set of possible states of $\rvm$
is $S_\rvm= S_{\rvx_{i_0}}$, and
the node matrix of $\rvm$ is
$P(\rvm=m|\rvx=x)=\delta(m,x_{i_0})$.

Let $\cbnet$ be a CB net for which we want to
obtain a q-embedding.
Our construction has two steps,
given by Fig.\ref{algo-q-emb}.

\begin{figure}[h]\begin{center}
\fbox{\parbox{5in}{

\MyCases{(Step 1) Add marginalizer nodes.}

More specifically, replace $\cbnet$ by a
modified  CB net $\cbnet_{mod}$ obtained
 as follows.
For each node $\rvx$ of $\cbnet$,
add a marginalizer node between $\rvx$
and every child of $\rvx$. If $\rvx$ has no
children, add a child (with
a delta function as probability
matrix) to it.

\MyCases{(Step 2) Replace node probability matrices
by their q-embeddings. Add ancilla nodes.}

More
specifically, replace $\cbnet_{mod}$
by a QB net $\qbnet$ obtained as follows.
For each
node of $\cbnet_{mod}$, except for
the marginalizer nodes that were
added in the previous step,
replace its node matrix by a
new node matrix which is a
q-embedding of the original node matrix.
Add a new node
for each ancilla index of
each new node matrix.
These new nodes will be called
{\bf ancilla nodes} (of either
the source or sink type) because
they correspond to ancilla indices.
}}
\caption{Algorithm for
constructing a q-embedding of a CB net}
\label{algo-q-emb}
\end{center}
\end{figure}

Consider a QB net $\qbnet$ which is a q-embedding
of a CB net $\cbnet$.
In the following sections,
we will label some nodes of
$\qbnet$
by an underlined group of symbols, such as
$\ax$, followed by an index enclosed
in angular brackets,
as in $\ax\av{4}$.
These indices
enclosed in angular
brackets will be called {\bf worldline indices}.
In $\qbnet$, a
sequence of
random variables such as
$\ax\av{1}, \ax\av{2}, \ldots \ax\av{n}$,
where node $\ax\av{1}$ is set to zero and
$\ax\av{n}$ is an external node of $\qbnet$,
will be called a {\bf worldline} of $ax$.
When using worldline
indices,
variables like $(x, \tilde{x},y, \tilde{y})$
that were used in describing
a q-embedding of a probability matrix
are replaced by:

\beq
\begin{array}{c}
(\tilde{x},x) = (x\av{i+1},x\av{i})\\
(y,\tilde{y}) = (y\av{j+1},y\av{j})\\
\end{array}
\;,
\eeq
for some integers $i$ and $j$.

Every QB net $\qbnet$ can be converted
into an equivalent quantum circuit $\qckt$.
To do so,
each worldline of $\qbnet$
is turned into the time history
of one or more qubits.
(The number of qubits in a worldline is $log_2$
of the number of possible states of node
$\av{1}$ of the worldline.)

In Ref.\cite{Tuc00} we
consider two CB nets
called
Two-Body
Scattering and Asia.
For each  of these CB nets $\cbnet$,
we perform the steps of Fig.\ref{algo-q-emb}
 to construct
a QB net $\qbnet$ that is
a q-embedding of $\cbnet$.
Ref.\cite{Tuc00}
gives $\cbnet$ and $\qbnet$, but
not an equivalent quantum circuit $\qckt$.
As examples and for the sake of completeness,
Appendix \ref{app-quan-ckts}
gives
quantum circuits
for Two-Body Scattering and Asia.

\section{Importance Sampling}
In this section, we will
propose a method for doing
importance sampling of a CB net on a {\it quantum}
computer. The traditional method for doing
importance sampling
of a CB net on a {\it classical} computer
is reviewed in Appendix \ref{app-imp-sam}.

Consider a CB net whose nodes
are labeled in topological order
by $(\rvx_1, \rvx_2, \ldots \rvx_{N_{nds}})\equiv \rvx$.
Assume that $E$ (evidence set)
and $H$ (hypotheses set)
are disjoint subsets of $Z_{1,N_{nds}}$,
with $Z_{1,N_{nds}}-E\cup H$ not
necessarily empty.
Let $X^c = Z_{1,N_{nds}}-X$ for any
$X\subset Z_{1,N_{nds}}$.
Assume that we are given
the prior evidence $(x)_E$, and
the number of samples $N_{sam}$
that we intend to collect.

Suppose $x'$ is an arbitrary point in $val(\rvx)$.
(We'll use the unprimed $x$,
as in $(x)_E$, to denote the evidence.)
The probability  matrices
associated with each node of our
CB net will be denoted by $P(x'_i|(x')_{pa(i)})$
for each $i\in Z_{1, N_{nds}}$.
In addition, we will assume we are given
{\bf sampling probability matrices},
associated with each node of our
CB net,  denoted by
$Q(x'_i|(x')_{pa(i)})$ for each $i\in Z_{1, N_{nds}}$.
In all cases, these sampling matrices are constrained to
satisfy

\beq
Q(x'_i|(x')_{pa(i)})= P(x'_i|(x')_{pa(i)})
\;\;\forall i\in E^c
\;.
\eeq
Two important special cases
of importance sampling are rejection sampling
and
likelihood weighted sampling.
For {\bf rejection sampling (RS)},

\beq
Q(x'_i|(x')_{pa(i)})= P(x'_i|(x')_{pa(i)})
\;\;\forall i\in Z_{1, N_{nds}}
\;.
\eeq
Hence, $Q(x') = P(x')$ for rejection sampling.
For {\bf likelihood weighted sampling (LWS) (a.k.a. likelihood weighting)},
\beq
Q(x'_i|(x')_{pa(i)})=
\left\{
\begin{array}{l}
P(x'_i|(x')_{pa(i)})
\;\;\forall i\in E^c\\
\delta(x_i, x_i')
\;\;\forall i\in E
\end{array}
\right.
\;.
\eeq
Hence, $Q(x') = \delta_{(x)_E}^{(x')_E}
\prod_{i\in E^c}P(x'_i|(x')_{pa(i)})$ for
likelihood weighted sampling.

\begin{figure}[h]\begin{center}
\fbox{\parbox{6in}{
\begin{verse}
For all $(x)_H$ $\{W[(x)_H]=0;\}$\\
$W_{tot}=0;$\\
For samples $k=1,2, \ldots,N_{sam}\{$\\
\hspace{2em}$L=1;$\\
\hspace{2em}For nodes $i=1,2, \ldots,N_{nds}\{$\\
\hspace{4em}\underline{ Use quantum computer to generate $\sam{x_i}{k}$ from $Q(x_i|(\sam{x}{k})_{pa(i)});$}\\
\hspace{4em}$//$Here, for LWS, $\sam{x_i}{k}==x_i$ when $i\in E$.\\
\hspace{4em}$//$$ pa(i)\subset Z_{1, i-1}$ so $(\sam{x}{k})_{pa(i)}$ known at this point.\\
\hspace{4em}if $i\in E$\{\\
\hspace{6em}if $\sam{x_i}{k}==x_i$\{\\
\hspace{8em}$L\;*=\;\frac{P(x_i|(\sam{x}{k})_{pa(i)})
}{Q(x_i|(\sam{x}{k})_{pa(i)})};$\\
\hspace{8em}$//$Here $\frac{P}{Q}= 1$ for RS and $\frac{P}{Q}=P$ for LWS.\\
\hspace{6em}\}else\{$//$LWS never enters here\\
\hspace{8em}go to next k;\\
\hspace{6em}$\}$\\
\hspace{4em}$\}$\\
\hspace{2em}$\}// i$ loop (nodes)\\
\hspace{2em}$W[(\sam{x}{k})_H]\;+=\;L;$ \\
\hspace{2em}$W_{tot}\;+=\;L;$\\
$\}// k$ loop (samples)\\
For all $(x)_H$ $\{P((x)_H|(x)_E)=\frac{W[(x)_H]}{W_{tot}};\}$\\
\end{verse}
}}
\caption{Algorithm for importance sampling of
 CB net on quantum computer.}
 \label{algo-imp-sam-quan}
\end{center}
\end{figure}

Under these assumptions, the
algorithm for importance sampling of a
CB net on a quantum computer
is given by Fig.\ref{algo-imp-sam-quan}
(expressed in pseudo-code,
pidgin C language). The only difference between
the classical algorithm of
Fig.\ref{algo-imp-sam-cla}
and the quantum one of
Fig.\ref{algo-imp-sam-quan}
is the underlined command.
In the quantum case, we use
a quantum computer instead of a classical one
to generate
$\sam{x_i}{k}\sim Q(x_i|(\sam{x}{k})_{pa(i)})$.
To do this, we can
find a q-embedding  of
$Q(x_i|(\sam{x}{k})_{pa(i)})$.
Starting from an a priori known pure
state of the parents $\ket{(\rvx)_{pa(i)}=
(\sam{x}{k})_{pa(i)}}$,
one applies the q-embedding
to it, and then finally one measures $x_i$.

For example, suppose for some $i$,
$x_i\in Bool$ and
$(x)_{pa(i)}=(x_1,x_0)\in Bool^2$.
Denote $x_i$ by $y$. Then a q-embedding
$A$ of $Q(x_i|(x)_{pa(i)})$ satisfies

\beq
A(y,\tilde{x}_1,\tilde{x}_0|
\tilde{y}=0,x_1,x_0)=\\
\sqrt{Q(y|x_1,x_0)}
\;\;
\delta_{x_1}^{\tilde{x}_1}
\delta_{x_0}^{\tilde{x}_0}
\;.
\eeq
Suppose that the a priori known
pure state of the parents is
$\ket{(\rvx)_{pa(i)}=(x'_1,x'_0)}$.
If we indicate
non-zero entries by a plus sign,

\beqa
A\ket{\tilde{y}=0,x'_1,x'_0} &=&
\begin{tabular}{r|r|r|r|r|r|}
     &{\tiny$(\tilde{y},x_1,x_2)=$}&&&&\\
     & {\tiny 000}& {\tiny 001} &{\tiny 010}& {\tiny 011} & $\cdots$ \\
\hline
{\tiny $(y,\tilde{x}_1,\tilde{x}_2)$= 000} &+& & &&$\cdots$ \\
{\tiny 001} &  &+&&&$\cdots$\\
{\tiny 010} &  &&+&&$\cdots$\\
{\tiny 011} &  &&&+&$\cdots$\\
{\tiny 100} &  +&&&&$\cdots$\\
{\tiny 101} &  &+&&&$\cdots$\\
{\tiny 110} &  &&+&&$\cdots$\\
{\tiny 111} &  &&&+&$\cdots$\\
\end{tabular}
\ket{\tilde{y}=0,x'_1,x'_0}
\\
&\rarrow&
\sum_{\vecb\in Bool^2}
e^{i\theta_\vecb\sigy}\otimes P_\vecb
\ket{\tilde{y}=0,x'_1,x'_0}\\
&=&
e^{i\theta_{x'_1x'_0}\sigy(2)}
\ket{\tilde{y}=0,x'_1,x'_0}
\;,
\label{eq-reduced-a}
\eeqa
for some $\theta_\vecb\in \RR$.
Here the right pointing
arrow means that the expression at the
origin of the arrow can
be extended to the expression
at the target of the arrow.
Note that,
according to Eq.(\ref{eq-reduced-a}),
it is not necessary
to perform all elementary operations
that constitute a decomposition
of $A$. We need only perform one
single-qubit rotation picked out
by the a priori known state of the parents.
The final step is to
measure $\rvy$ on the state of Eq.(\ref{eq-reduced-a}),
without measuring $\rvx_1$ and $\rvx_0$.

If $\rvx_i$ or one of its parent nodes
 has more than two possible states,
then (see Section \ref{sec-q-emb-mats})
we can still represent
the q-embedding $A$
as a multiplexor of a chain
of multiplexors.
This will give for
$A\ket{\tilde{x_i}=0,(\sam{x}{k})_{pa(i)}}$
a chain of multiplexors
acting on
$\ket{\tilde{x_i}=0,(\sam{x}{k})_{pa(i)}}$.
The final step is to
measure $\rvx_i$,
without measuring $(\rvx)_{pa(i)}$.

Note that since nodes $\rvx_j$
for $j\in E$
are fixed, we may treat them as if each
had only one possible state. This will
reduce the size of the probability
matrix $P(x_i|(x)_{pa(i)})$,
and of its q-embedding $A$.

\section{Markov Chain Monte Carlo}
\subsection{Gibbs Sampling}

In this section, we will
propose a method for doing
Gibbs sampling of a CB net on a {\it quantum}
computer. The traditional method for doing
Gibbs sampling
of a CB net on a {\it classical} computer
is reviewed in Appendix \ref{app-gibbs-sam}.

Consider a Markov chain
$\rvx^0\rarrow\rvx^1\rarrow\rvx^2\ldots \rarrow\rvx^T$.
Let
$\rvx^t = (\rvx_1^t, \rvx_2^t,\ldots,\rvx^t_{N_{nds}})$
for each time $t$
represent a separate copy of a CB net
with nodes $\rvx_1^t, \rvx_2^t,\ldots,\rvx^t_{N_{nds}}$,
and probability matrices
$P(x_i^t| (x^t)_{pa(i)})$.
(The nodes of the CB net
$\rvx^t$ are
 not necessarily in topological order.)
Assume that $E$ (evidence set)
and $H$ (hypotheses set)
are disjoint subsets of $Z_{1,N_{nds}}$,
with $Z_{1,N_{nds}}-E\cup H$ not
necessarily empty.
Let $X^c = Z_{1,N_{nds}}-X$ for any
$X\subset Z_{1,N_{nds}}$.
Assume that we are given
the prior evidence $(x)_E$.
All probabilities in this
section about the Gibbs algorithm
will be conditioned implicitly on
$(\rvx^t)_E = (x)_E$ for all $t$. The
Gibbs algorithm is designed to respect this constraint,
by never changing the value of $(\rvx^t)_E$
after it is initially set.

Suppose that we wish to
sweep through all nodes of graph $\rvx^t$,
in a fixed deterministic order,
repeating this all-nodes-sweep $N_{gra}$ times.
$t$ will change by one every time
one node is visited. Thus,
the last time $T$ of the Markov chain
will be $N_{nds}N_{gra}$.
Suppose we wish to sweep through $\beta$
copies of the graph $\rvx^t$ before
performing each measurement.
Assume that we are given
the burn time $t_{burn}$ ($0<<t_{burn}<<T$).

\begin{figure}[h]\begin{center}
\fbox{\parbox{5.5in}{
\begin{verse}
For all $(x)_H$ $\{W[(x)_H]=0;\}$\\
$W_{tot}=0;$\\
Initialize $\rvx^0$ to some value $x^0$, subject to $(x^0)_E=(x)_E;$\\
$t=0;$\\
For graphs $g=\beta,2\beta,3\beta, \ldots,N_{gra}\{$\\
\hspace{2em}\underline{Use quantum computer to generate $x^{t+\beta N_{nds}}\sim P(x^{t+\beta N_{nds}}|x^t);$}\\
\hspace{2em}\underline{$t += \beta N_{nds};$}\\
\hspace{2em}if $t> t_{burn}$\{$//$ $0<<t_{burn}<< N_{nds}N_{gra}$\\
\hspace{4em}$W[(x^{t})_H]++;$ \\
\hspace{4em}$W_{tot}++;$\\
\hspace{2em}$\}$\\
$\}// g$ loop (graphs)\\
For all $(x)_H$ $\{P((x)_H|(x)_E)=\frac{W[(x)_H]}{W_{tot}};\}$\\
\end{verse}
}}
\caption{Algorithm for Gibbs sampling of
 CB net on quantum computer.
 Nodes of CB net $\rvx^t$
 are visited in a fixed deterministic order.}
 \label{algo-gibbs-sam-quan}
\end{center}
\end{figure}

Under these assumptions, the
algorithm for Gibbs sampling of a
CB net on a quantum computer
is given by Fig.\ref{algo-gibbs-sam-quan}
(expressed in pseudo-code,
pidgin C language).
When $\beta=1$, the only difference between
the classical algorithm of
Fig.\ref{algo-gibbs-sam-cla-det}
and the quantum one of
Fig.\ref{algo-gibbs-sam-quan}
is that the two underlined commands
in the quantum algorithm replace
 the $i$ loop (over nodes)
in the classical one.
In the quantum case, we use
a quantum computer
to generate
$x^{t+\beta N_{nds}}\sim P(x^{t+\beta N_{nds}}|x^t)$.
To do this, we
find a CB net
that generates
$P(x^{t+\beta N_{nds}}|x^t)$
and then we find a q-embedding  of
that CB net.

\begin{figure}[h]
    \begin{center}
    \epsfig{file=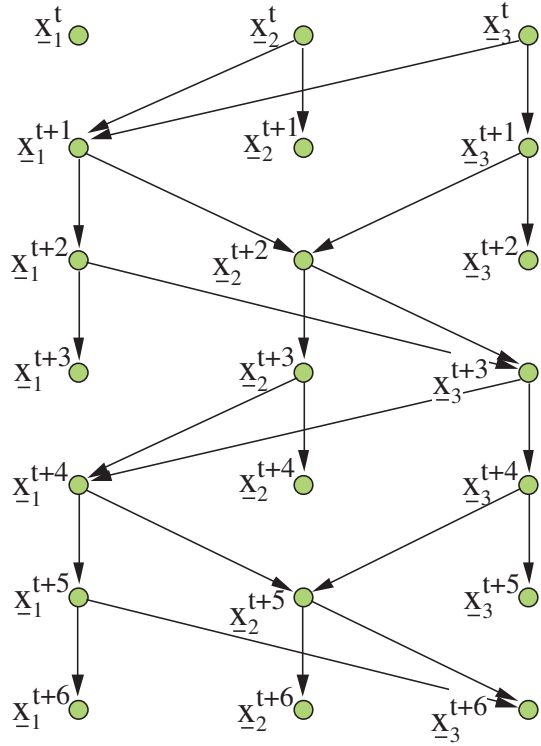, height=4in}
    \caption{CB net for Gibbs sampling algorithm, for
    a CB net $\rvx^t$ with 3 nodes,
    sweeping through the nodes
    of $\rvx^t$ in a fixed deterministic order,
    repeating this all-nodes-sweep twice. }
    \label{fig-gibbs-cbnet}
    \end{center}
\end{figure}

As an example, suppose
$N_{nds}=3$ and $\beta=2$.
Fig.\ref{fig-gibbs-cbnet}
shows a CB net
that will generate
$P(x^{t+\beta N_{nds}}|x^t)$ provided that
we set

\beq
P(x^t_i) =\delta(x^t_i,(x^t_i)_{prev})
\;
\label{eq-prob-first-t}
\eeq
for $i\in Z_{1,3}$.
Here $(x^t_i)_{prev}$ are the values
of $x^t_i$ obtained previously in the algorithm.

Note that Fig.\ref{fig-gibbs-cbnet} is
divided into 7 ``time slices"
$t+j$ for $j\in Z_{0,6}$.
Eq.(\ref{eq-prob-first-t}) gives
the probability matrices associated
with the nodes of the first time slice.
The probability matrices
associated with the nodes of the other 6
time slices
are
as follows. For $j\in Z_{0,5}$,

\begin{subequations}
\beq
P(x_{1\oplus j}^{t+1+j}|
x_{2\oplus j}^{t+j}, x_{3\oplus j}^{t+j})=
P_{\rvx_{1\oplus j}|
\rvx_{2\oplus j}, \rvx_{3\oplus j}}
(x_{1\oplus j}^{t+1+j}|
x_{2\oplus j}^{t+j}, x_{3\oplus j}^{t+j})
\;,
\eeq

\beq
P(x_{2\oplus j}^{t+1+j}|x_{2\oplus j}^{t+j}) = \delta(x_{2\oplus j}^{t+1+j},x_{2\oplus j}^{t+j})
\;,
\eeq

\beq
P(x_{3\oplus j}^{t+1+j}|x_{3\oplus j}^{t+j}) = \delta(x_{3\oplus j}^{t+1+j},x_{3\oplus j}^{t+j})
\;,
\eeq
\end{subequations}
where  $\oplus$
denotes addition mod 3
with 3 and 0 identified.\footnote{
In C language,
${\tt x\oplus y =
((x+y)\%3==0?3:(x+y)\%3)}$,
where $x,y$ are non-negative integers.
}
Here
$P_{\rvx_{1\oplus j}|
\rvx_{2\oplus j}, \rvx_{3\oplus j}}$
is a node probability of the
CB net $\rvx^t$.
We can replace
$P_{\rvx_{1\oplus j}|
\rvx_{2\oplus j}, \rvx_{3\oplus j}}$
by
$P_{\rvx_{1\oplus j}|
(\rvx)_{MB(1\oplus j)}}$
if the Markov blanket of
$\rvx_{1\oplus j}$
does not include all
nodes in $(\rvx)_{\{1\oplus j\}^c}$.
For example, when $j=0$, we get the
conditional probabilities
of the nodes of the first
time slice:

\begin{subequations}
\beq
P(x_{1}^{t+1}|
x_{2}^{t}, x_{3}^{t})=
P_{\rvx_{1}|
\rvx_{2}, \rvx_{3}}
(x_{1}^{t+1}|
x_{2}^{t}, x_{3}^{t})
\;,
\eeq

\beq
P(x_{2}^{t+1}|x_{2}^{t}) = \delta(x_{2}^{t+1},x_{2}^{t})
\;,
\eeq

\beq
P(x_{3}^{t+1}|x_{3}^{t}) = \delta(x_{3}^{t+1},x_{3}^{t})
\;.
\eeq
\end{subequations}
Here
$P_{\rvx_{1}|
\rvx_{2}, \rvx_{3}}$
is a node probability of the
CB net $\rvx^t$.
We can replace
$P_{\rvx_{1}|
\rvx_{2}, \rvx_{3}}$
by
$P_{\rvx_{1}|
(\rvx)_{MB(1)}}$
if the Markov blanket of
$\rvx_1$
does not include all
nodes in $(\rvx)_{\{1\}^c}$.

The full probability
distribution for the net of
Fig.\ref{fig-gibbs-cbnet}
is given by:

\beq
P(x^t, x^{t+1}, x^{t+2},x^{t+3},x^{t+4},x^{t+5},x^{t+6})=
\left\{\vcenter{\xymatrix{
P(x_1^t)&P(x_2^t)\ar@{-}[d]&P(x_3^t)\ar@{-}[d]\\
P(x_1^{t+1}\ar@{-}[drr]&|x_2^{t+1}&,x_3^{t+1})\ar@{-}[dl]\\
P(x_2^{t+2}\ar@{-}[drr]&|x_3^{t+2}&,x_1^{t+2})\ar@{-}[dl]\\
P(x_3^{t+3}\ar@{-}[drr]&|x_1^{t+3}&,x_2^{t+3})\ar@{-}[dl]\\
P(x_1^{t+4}\ar@{-}[drr]&|x_2^{t+4}&,x_3^{t+4})\ar@{-}[dl]\\
P(x_2^{t+5}\ar@{-}[drr]&|x_3^{t+5}&,x_1^{t+5})\ar@{-}[dl]\\
P(x_3^{t+6}&|x_1^{t+6}&,x_2^{t+6})
}}
\right\}
\;,
\eeq
where each of the slanted lines represents
 a Kronecker delta function equating
the two variables at the two ends of the line.
Summing the full probability distribution over all nodes except the
final ones $\rvx^{t+6}$ yields:

\beq
P(x^{t+6}) =
\sum_{x_1^{t},x_2^{t+1},x_3^{t+2}}
\;\;\;\;
\sum_{x_1^{t+3},x_2^{t+4},x_3^{t+5}}
\left\{
\begin{array}{l}
P(x_1^t)P(x_2^{t+1})P(x_3^{t+2})\\
P(x_1^{t+3}|x_2^{t+1},x_3^{t+2})\\
P(x_2^{t+4}|x_3^{t+2},x_1^{t+3})\\
P(x_3^{t+5}|x_1^{t+3},x_2^{t+4})\\
P(x_1^{t+6}|x_2^{t+4},x_3^{t+5})\\
P(x_2^{t+6}|x_3^{t+5},x_1^{t+6})\\
P(x_3^{t+6}|x_1^{t+6},x_2^{t+6})
\end{array}
\right\}
\;.
\label{eq-ptplussix-ugly}
\eeq
It is convenient to make the
following change of notation:

\beq
\begin{array}{l}
(x_1^{t},x_2^{t+1},x_3^{t+2})\rarrow
(X_1^{t},X_2^{t},X_3^{t})
\\
(x_1^{t+3},x_2^{t+4},x_3^{t+5})\rarrow
(X_1^{t+3},X_2^{t+3},X_3^{t+3})
\\
x^{t+6}\rarrow X^{t+6}
\end{array}
\;
\eeq
Described more succinctly, what we
are
doing is replacing $x$ by $X$ and
$t +j$ by $t + (j/3)3$ for
$j\in Z_{0,6}$, where the
division by 3 is ``integer division",
with no remainder, an operation available
in most computer languages.
In the new notation,
Eq.(\ref{eq-ptplussix-ugly}) simplifies to

\beq
P(x^{t+6}) =
\sum_{X^{t}}
\;\;\;\;
\sum_{X^{t+3}}
\left\{
\begin{array}{l}
P(X^t)\\
P(X_1^{t+3}|X_2^{t},X_3^{t})\\
P(X_2^{t+3}|X_3^{t},X_1^{t+3})\\
P(X_3^{t+3}|X_1^{t+3},X_2^{t+3})\\
P(X_1^{t+6}|X_2^{t+3},X_3^{t+3})\\
P(X_2^{t+6}|X_3^{t+3},X_1^{t+6})\\
P(X_3^{t+6}|X_1^{t+6},X_2^{t+6})
\end{array}
\right\}
\;,
\eeq
or, equivalently,

\beq
P(x^{t+6})=
\sum_{X^t, X^{t+3}}
P(X^t)
\prod_{j=1,2}
\left\{
\begin{array}{l}
P(X_1^{t+3j}|X_2^{t+3(j-1)},X_3^{t+3(j-1)})\\
P(X_2^{t+3j}|X_3^{t+3(j-1)},X_1^{t+3j})\\
P(X_3^{t+3j}|X_1^{t+3j},X_2^{t+3j})
\end{array}
\right\}
\;.
\label{eq-prob-xsix}
\eeq

\clearpage
\begin{figure}[h]
    \begin{center}
    \epsfig{file=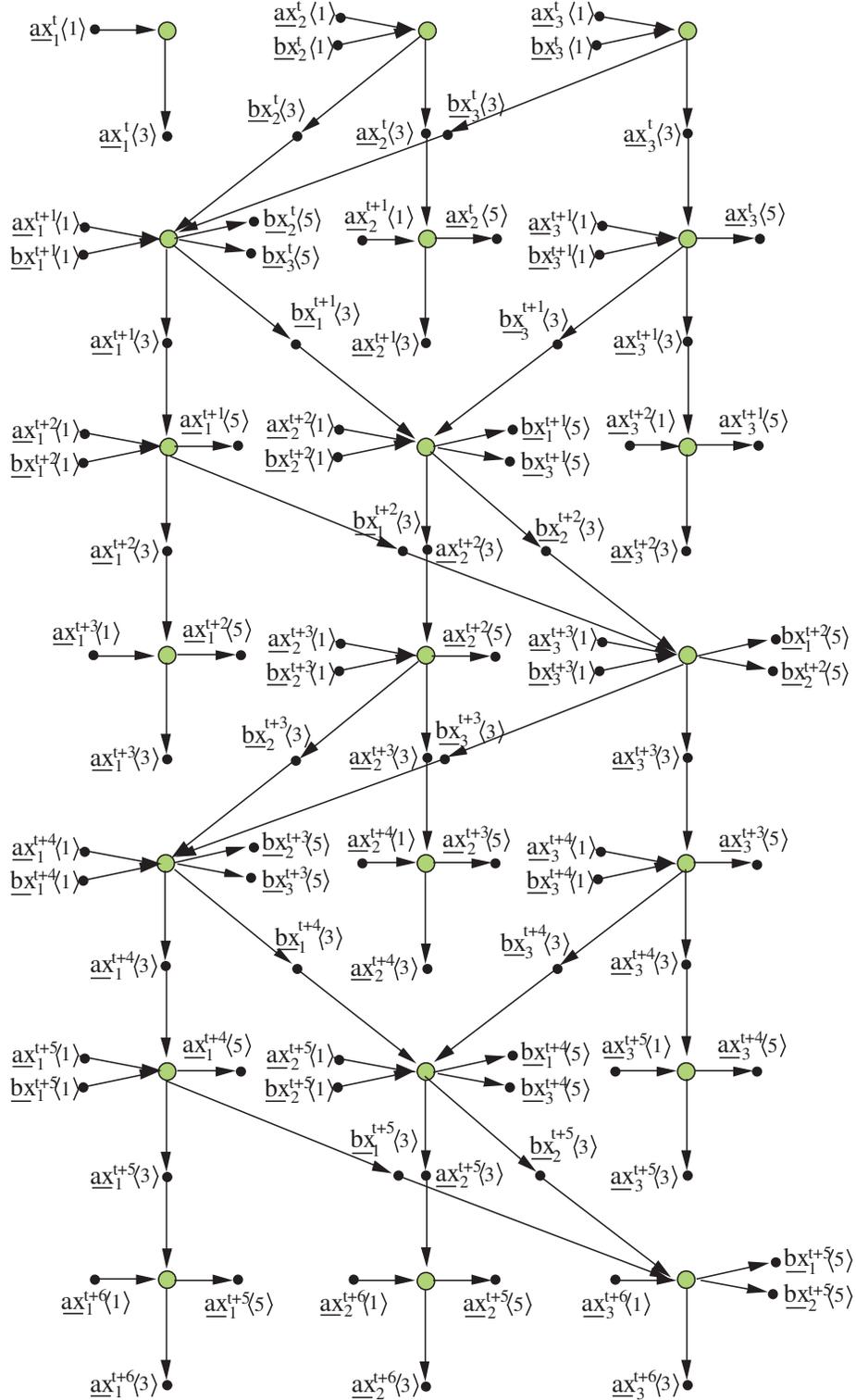, height=8in}
    \caption{QB net that is a q-embedding of the
     CB net of
    Fig.\ref{fig-gibbs-cbnet}.}
    \label{fig-gibbs-qbnet}
    \end{center}
\end{figure}
\clearpage

\begin{figure}[h!]
    \begin{center}
    \epsfig{file=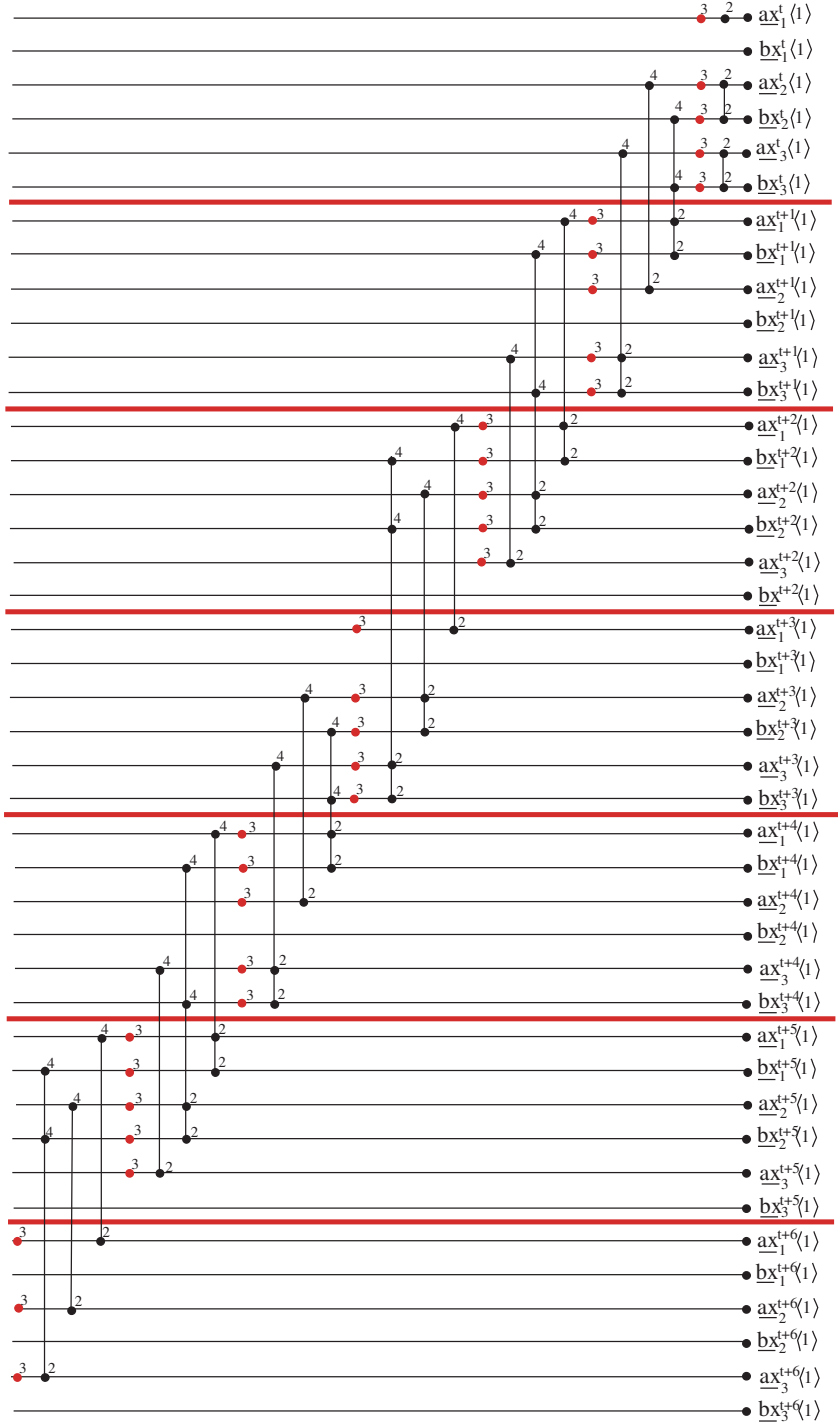, height=7.5in}
    \caption{Quantum circuit for QB net of
    Fig.\ref{fig-gibbs-qbnet}}
    \label{fig-gibbs-ckt}
    \end{center}
\end{figure}

Following the
steps of Fig.\ref{algo-q-emb}, we
obtain the QB net
Fig.\ref{fig-gibbs-qbnet},
a q-embedding
 of the CB net of Fig.\ref{fig-gibbs-cbnet}.
 In Fig.\ref{fig-gibbs-qbnet},
due to lack of space,
labels for the original
(light-colored, not black) nodes
have been omitted. These omitted
labels
can be re-constituted as follows. If
the original node has incoming
arrows $\rvz_1, \rvz_2, \ldots, \rvz_n$,
then the node is labeled by
the n-tuple $(\rvz_1, \rvz_2, \ldots, \rvz_n)$.

From the QB net of
 Fig.\ref{fig-gibbs-qbnet},
 one easily obtains
 the equivalent quantum circuit
 shown in Fig.\ref{fig-gibbs-ckt}.
Some simple observations about
this circuit are:
\begin{itemize}

\item
The first  time  slice $t$ uses only 4 qubits,
the intermediate ones use 5 qubits,
and the final one $t+6$ uses 3 qubits.
\item
Once the ``column of $\av{3}$'s" is
reached midway into time slice $t+j$
for $j\in Z_{1,5}$,
the qubits of the previous time
slice can be recycled (used again).
Also, in each time slice, at least one
of the six qubits is never used. Thus,
we only need $5+5=10$ qubits for $N_{nds}=3$,
or $2(2N_{nds}-1)$ in general.
\item
All worldlines except those of
the first time slice start at $\ket{0}$.
In the first time slice,
$ax^t_i\av{1}$ and $bx^t_i\av{1}$ start
at $\ket{(x^t_i)_{prev}}$ for $i=1,2,3$.
\item
Nodes $\av{3}$ may be omitted since
they do not change the state of their qubit.

\end{itemize}

Let $\Gamma_{ext}$ be the
set of all external nodes of the
QB net of Fig.\ref{fig-gibbs-qbnet},
and let $\Gamma_{int}$ be the set of
all other nodes. Let
$A$ be the full amplitude of
the QB net. Then

\beq
P(ax^{t+6}\av{3}|(x^t)_{prev})=
\sum_{\Gamma_{ext}-\{ax^{t+6}\av{3}\}}
\left|\sum_{\Gamma_{int}}
A\right|^2
\;.
\eeq
By construction,
the probability $P(ax^{t+6}\av{3}|(x^t)_{prev})$
and the probability
$P(x^{t+6}|(x^t)_{prev})$
of Eq.(\ref{eq-prob-xsix})
should be equal, if we equate
$ax^{t+6}\av{3}$ and $x^{t+6}$.

Next, we will give the node
amplitudes for the nodes of
quantum circuit
Fig.\ref{fig-gibbs-ckt}.
We will give node
amplitudes only when those
amplitudes are non-trivial
(i.e.,
not equal to just a delta function,
like they are for the $\av{1}$ and $\av{3}$ nodes).
We will give compilations
for these non-trivial node amplitudes assuming
$val(\rva)\in Bool$ for all
nodes $\rva$ of the CB net $\rvx^t$.
If some node has more than
two possible values, then,
we increase the number of
values of that node to a power of
two. Compilation of
node amplitudes in this case is slightly
more complicated than when
all nodes
have only
two possible values,
but it can still be done using
the multiplexor techniques of
Section \ref{sec-q-emb}.

{\bf Nontrivial nodes in initial time slice $t$:}

\begin{enumerate}
\item Nodes
$(\ax_2^{t}\av{2},\bx_2^t\av{2})$ and
$(\ax_3^{t}\av{2},\bx_3^t\av{2})$.
These two nodes are analogous. Consider
the first one for definiteness. One has

\beq
\begin{array}{l}
A(ax_2^{t}\av{2},bx_2^{t}\av{2}|
ax_2^{t}\av{1}=(x^t_2)_{prev},bx_2^{t}\av{1}=(x^t_2)_{prev})=
\\ \;\;
\Theta(
ax_2^{t}\av{2}=ax_2^{t}\av{1}=(x^t_2)_{prev})
\Theta(
bx_2^{t}\av{2}=bx_2^{t}\av{1}=(x^t_2)_{prev})
\end{array}
\;.
\eeq
One can extend this $A$
to the identity matrix,
if it acts on
{\tiny $\ket{\ax_2^t\av{1}=(x^t_2)_{prev}}
\ket{\bx_2^t\av{1}=(x^t_2)_{prev}}$}.

\item Node
$\ax_1^{t}\av{2}$. One has

\beq
\begin{array}{l}
A(ax_1^{t}\av{2}|ax_1^{t}\av{1}=(x^t_1)_{prev})=
\\ \;\;
\Theta(
ax_1^{t}\av{2}=ax_1^{t}\av{1}=(x^t_1)_{prev})
\end{array}
\;.
\eeq
One can extend this $A$
to the identity matrix,
if it acts on $\ket{\ax_1^t\av{1}=(x^t_1)_{prev}}$.

{\bf Nontrivial nodes straddling
time slices $t+j$ and $t+j+1$ for $j\in Z_{0,4}$:}
\end{enumerate}

\begin{enumerate}

\item
Nodes
$(\bx_{f(j)}^{t+1+j}\av{2},
\ax_{f(j)}^{t+1+j}\av{2},
\bx_{f(j)\oplus 2}^{t+j}\av{4},
\bx_{f(j)\oplus 1}^{t+j}\av{4}
)$
where $\begin{array}{c|ccccc}
j&0&1&2&3&4\\
\hline
f(j)&1&2&3&1&2\\
\end{array}$, and the $\oplus$
denotes mod 3 addition
with 3 and 0 identified.
 These five nodes
are analogous. Consider $j=0$ for definiteness.

For node $(\bx_1^{t+1}\av{2},
\ax_1^{t+1}\av{2},
\bx_3^{t}\av{4}),
\bx_2^{t}\av{4})$, one has

\beq
\begin{array}{l}
A(
bx_1^{t+1}\av{2},
ax_1^{t+1}\av{2},
bx_3^{t}\av{4},
bx_2^{t}\av{4}
|
bx_1^{t+1}\av{1}=0,
ax_1^{t+1}\av{1}=0,
bx_3^{t}\av{3},
bx_2^{t}\av{3})=
\\ \;\;
\sqrt{
P_{\rvx_1|\rvx_3,\rvx_2}
(ax_1^{t+1}\av{2}|
bx_3^{t}\av{3},
bx_2^{t}\av{3}
)
}\;\;
\delta_{ax_1^{t+1}\av{2}}^{bx_1^{t+1}\av{2}}
\delta_{bx_3^{t}\av{3}}^{bx_3^{t}\av{4}}
\delta_{bx_2^{t}\av{3}}^{bx_2^{t}\av{4}}
\end{array}
\;.
\eeq
If we indicate non-zero entries by a plus sign,

\beqa
A &=&
\begin{tabular}{r|r|r|r|r|r|}
     & {\tiny 0000}& {\tiny 0001} &{\tiny 0010}& {\tiny 0011} & $\cdots$ \\
\hline
{\tiny $(bx_1^{t+1},ax_1^{t+1},bx_3^{t},bx_2^{t})$= 0000} &+& & &&$\cdots$\\
{\tiny 0001} &  &+&&&$\cdots$\\
{\tiny 0010} &  &&+&&$\cdots$\\
{\tiny 0011} &  &&&+&$\cdots$\\
{\tiny 0100} &  &&&&$\cdots$\\
{\tiny 0101} &  &&&&$\cdots$\\
{\tiny 0110} &  &&&&$\cdots$\\
{\tiny 0111} &  &&&&$\cdots$\\
{\tiny 1001} &  &&&&$\cdots$\\
{\tiny 1010} &  &&&&$\cdots$\\
{\tiny 1011} &  &&&&$\cdots$\\
{\tiny 1100} &  +&&&&$\cdots$\\
{\tiny 1101} &  &+&&&$\cdots$\\
{\tiny 1110} &  &&+&&$\cdots$\\
{\tiny 1111} &  &&&+&$\cdots$\\
\end{tabular}
\\
&\rarrow&
\sigx(3)^{n(2)}
I_2\otimes \sum_{\vecb\in Bool^2}e^{i\theta_\vecb \sigy}
\otimes P_\vecb\\
&=&
\sigx(3)^{n(2)}
\sum_{\vecb\in Bool^2}e^{i\theta_\vecb \sigy(2)}P_\vecb(1,0)
\;,
\eeqa
for some $\theta_\vecb\in \RR$.
This choice of $A$ can be compiled using
multiplexor methods.
\item
Nodes $(\ax_{f(j)}^{t+1+j}\av{2},\ax_{f(j)}^{t+j}\av{4})$
where $\begin{array}{c|ccccc}
j&0&1&2&3&4\\
\hline
f(j)&2&3&1&2&3
\end{array}$.
These five nodes
are analogous. Consider $j=0$ for definiteness.

For node
$(\ax_2^{t+1}\av{2},\ax_2^{t}\av{4})$, one has

\beq
\begin{array}{l}
A(ax_2^{t+1}\av{2},ax_2^{t}\av{4}|
ax_2^{t+1}\av{1}=0,ax_2^{t}\av{3})
=\\ \;\;
\Theta(
ax_2^{t+1}\av{2}=
ax_2^{t}\av{4}=
ax_2^{t}\av{3})
\end{array}
\;.
\eeq
Thus,

\beqa
A &=&
\begin{tabular}{r|r|r|r|}
&{\tiny $(ax_2^{t+1}\av{1},ax_2^{t}\av{3})=$}&&\\
     & {\tiny 00} & {\tiny 01} & $\cdots$ \\
\hline
{\tiny $(ax_2^{t+1}\av{2},ax_2^{t}\av{4})$= 00} &1&0& $\cdots$  \\
{\tiny 01} &  0&0&$\cdots$ \\
{\tiny 10} &  0&0&$\cdots$ \\
{\tiny 11} &  0&1&$\cdots$
\end{tabular}
\\
&\rarrow&
\sigx(1)^{n(0)} I_2^{\otimes 2}
\;.
\eeqa

This operation is unnecessary except
at the end, between time slices
$t+5$ and $t+6$.

\item Nodes
$(\bx_{f(j)}^{t+1+j}\av{2},\ax_{f(j)}^{t+1+j}\av{2},\ax_{f(j)}^{t+j}\av{4})$
where $\begin{array}{c|ccccc}
j&0&1&2&3&4\\
\hline
f(j)&3&1&2&3&1
\end{array}$. These five nodes
are analogous. Consider $j=0$ for definiteness.

For node
$(\bx_3^{t+1}\av{2},\ax_3^{t+1}\av{2},\ax_3^{t}\av{4})$,
one has

\beq
\begin{array}{l}
A(bx_3^{t+1}\av{2},ax_3^{t+1}\av{2},ax_3^{t}\av{4}|
bx_3^{t+1}\av{1}=0,ax_3^{t+1}\av{1}=0,ax_3^{t}\av{3})=
\\ \;\;
\Theta(
bx_3^{t+1}\av{2}=
ax_3^{t+1}\av{2}=
ax_3^{t}\av{4}=
ax_3^{t}\av{3})
\end{array}
\;.
\eeq
Thus,

\beqa
A &=&
\begin{tabular}{r|r|r|r|}
&{\tiny $(bx_3^{t+1}\av{1},ax_3^{t+1}\av{1},ax_3^{t}\av{3})=$}&&\\
     & {\tiny 000}& {\tiny 001} & $\cdots$ \\
\hline
{\tiny $(bx_3^{t+1}\av{2},ax_3^{t+1}\av{2},ax_3^{t}\av{4})$= 000} &1& 0 &$\cdots$ \\
{\tiny 001} &  0&0&$\cdots$\\
{\tiny 010} &  0&0&$\cdots$\\
{\tiny 011} &  0&0&$\cdots$\\
{\tiny 100} &  0&0&$\cdots$\\
{\tiny 101} &  0&0&$\cdots$\\
{\tiny 110} &  0&0&$\cdots$\\
{\tiny 111} &  0&1&$\cdots$\\
\end{tabular}
\\
&\rarrow&
[\sigx(2)\sigx(1)]^{n(0)} I_2^{\otimes 3}
\;.
\eeqa

\end{enumerate}

{\bf Nontrivial nodes straddling
time slices $t+5$ and $t+6$:}

\begin{enumerate}
\item Nodes
$(\ax_1^{t+6}\av{2},
\ax_1^{t+5}\av{4}
)$ and
$(\ax_2^{t+6}\av{2},
\ax_2^{t+5}\av{4}
)$. These two nodes are analogous. Consider
the first one for definiteness. One has

\beq
\begin{array}{l}
A(ax_1^{t+6}\av{2},
ax_1^{t+5}\av{4}|
ax_1^{t+6}\av{1}=0,
ax_1^{t+5}\av{3})=\\
\;\;
\Theta(
ax_1^{t+6}\av{2}=
ax_1^{t+5}\av{4}=
ax_1^{t+5}\av{3}
)
\end{array}
\;.
\eeq
Thus,

\beqa
A &=&
\begin{tabular}{r|r|r|r|}
&{\tiny $(ax_1^{t+6}\av{1},
ax_1^{t+5}\av{3})=$}&&\\
     & {\tiny 00} & {\tiny 01} & $\cdots$ \\
\hline
{\tiny $(ax_1^{t+6}\av{2},
ax_1^{t+5}\av{4})$= 00} &1&0&$\cdots$  \\
{\tiny 01} &  0&0&$\cdots$\\
{\tiny 10} &  0&0&$\cdots$\\
{\tiny 11} &  0&1&$\cdots$
\end{tabular}
\\
&\rarrow&
\sigx(1)^{n(0)} I_2^{\otimes 2}
\;.
\eeqa

\item Node
$(\ax_3^{t+6}\av{2},
\bx_2^{t+5}\av{4},
\bx_1^{t+5}\av{4}
)$. One has

\beq
\begin{array}{l}
A(
ax_3^{t+6}\av{2},
bx_2^{t+5}\av{4},
bx_1^{t+5}\av{4}|
ax_3^{t+6}\av{1}=0,
bx_2^{t+5}\av{3},
bx_1^{t+5}\av{3})=\\
\;\;
\sqrt{P_{\rvx_3|\rvx_2,\rvx_1}
(ax_3^{t+6}\av{2}|
bx_2^{t+5}\av{3},
bx_1^{t+5}\av{3}
}\;\;
\delta_{bx_2^{t+5}\av{3}}^{bx_2^{t+5}\av{4}}
\delta_{bx_1^{t+5}\av{3}}^{bx_1^{t+5}\av{4}}
\end{array}
\;.
\eeq
If we indicate
non-zero entries by a plus sign,

\beqa
A &=&
\begin{tabular}{r|r|r|r|r|r|}
     & {\tiny 000}& {\tiny 001} &{\tiny 010}& {\tiny 011} & $\cdots$ \\
\hline
{\tiny $(ax_3^{t+6},
bx_2^{t+5},
bx_1^{t+5})$= 000} &+& & &&$\cdots$ \\
{\tiny 001} &  &+&&&$\cdots$\\
{\tiny 010} &  &&+&&$\cdots$\\
{\tiny 011} &  &&&+&$\cdots$\\
{\tiny 100} &  +&&&&$\cdots$\\
{\tiny 101} &  &+&&&$\cdots$\\
{\tiny 110} &  &&+&&$\cdots$\\
{\tiny 111} &  &&&+&$\cdots$\\
\end{tabular}
\\
&\rarrow&
\sum_{\vecb\in Bool^2}
e^{i\theta_\vecb\sigy}\otimes P_\vecb\\
&=&
\sum_{\vecb\in Bool^2}
e^{i\theta_\vecb\sigy(2)}P_\vecb(1,0)
\;,
\eeqa
for some $\theta_\vecb\in \RR$.
This choice of $A$ can be compiled using
multiplexor methods.

\end{enumerate}

\subsection{Metropolis-Hastings Sampling}

In this section, we will
propose a method for doing
Metropolis-Hastings sampling of a CB net on a {\it quantum}
computer. The traditional method for doing
Metropolis-Hastings sampling
of a CB net on a {\it classical} computer
is reviewed in Appendix \ref{app-met-has-sam}.

Compare Eq.(\ref{eq-gibbs-p-trans-i})
for the Gibbs algorithm with
Eq.(\ref{eq-met-has-p-trans-i})
for the Metropolis-Hastings algorithm.
From this comparison we conclude
that if in the Gibbs algorithm of
Fig.\ref{algo-gibbs-sam-quan}, we replace
$P(x_i^{t+1}|(x^t)_{MB(i)})$ for $i\in E^c$
by
the following, we will
be doing Metropolis-Hastings.

\beqa
P(x_i^{t+1}|(x^t)_{MB(i)}) &\rarrow&
\qbar_i(x_i^{t+1}|x^t)
+
\delta^{x_i^t}_{x_i^{t+1}}
[1-\sum_{y_i}\qbar_i(y_i|x^t)]
\\
&=&
\left\{
\begin{array}{l}
\Theta(x^t_i\neq x_i^{t+1})
\qbar_i(x_i^{t+1}|x^t)\\
+\\
\Theta(x^t_i= x_i^{t+1})
[1-\sum_{y_i: y_i\neq x^t_i}\qbar_i(y_i|x^t)]
\end{array}
\right\}
\;.
\eeqa

\begin{appendix}

\section{Appendix:
Importance Sampling \\of CB Nets
on a Classical Computer}\label{app-imp-sam}

In this Appendix, we review
the importance sampling algorithm
for CB nets
on a classical computer.

Consider a CB net whose nodes
are labeled in topological order
by $(\rvx_1, \rvx_2, \ldots \rvx_{N_{nds}})\equiv \rvx$.
Assume that $E$ (evidence set)
and $H$ (hypotheses set)
are disjoint subsets of $Z_{1,N_{nds}}$,
with $Z_{1,N_{nds}}-E\cup H$ not
necessarily empty.
Let $X^c = Z_{1,N_{nds}}-X$ for any
$X\subset Z_{1,N_{nds}}$.
Assume that we are given
the prior evidence $(x)_E$, and
the number of samples $N_{sam}$
that we intend to collect.

Suppose $x'$ is an arbitrary point in $val(\rvx)$.
(We'll use the unprimed $x$,
as in $(x)_E$, to denote the evidence.)
The probability  matrices
associated with each node of our
CB net will be denoted by $P(x'_i|(x')_{pa(i)})$
for each $i\in Z_{1, N_{nds}}$.
In addition, we will assume we are given
{\bf sampling
 probability matrices},
associated with each node of our
CB net,  denoted by
$Q(x'_i|(x')_{pa(i)})$ for each $i\in Z_{1, N_{nds}}$.
In all cases, these sampling matrices are constrained to
satisfy

\beq
Q(x'_i|(x')_{pa(i)})= P(x'_i|(x')_{pa(i)})
\;\;\forall i\in E^c
\;.
\eeq
Two important special cases
of importance sampling are rejection sampling
and
likelihood weighted sampling.
For {\bf rejection sampling (RS)},

\beq
Q(x'_i|(x')_{pa(i)})= P(x'_i|(x')_{pa(i)})
\;\;\forall i\in Z_{1, N_{nds}}
\;.
\eeq
Hence, $Q(x') = P(x')$ for rejection sampling.
For {\bf likelihood weighted sampling (LWS) (a.k.a. likelihood weighting)},
\beq
Q(x'_i|(x')_{pa(i)})=
\left\{
\begin{array}{l}
P(x'_i|(x')_{pa(i)})
\;\;\forall i\in E^c\\
\delta(x_i, x_i')
\;\;\forall i\in E
\end{array}
\right.
\;.
\eeq
Hence, $Q(x') = \delta_{(x)_E}^{(x')_E}
\prod_{i\in E^c}P(x'_i|(x')_{pa(i)})$ for
likelihood weighted sampling.

Under these assumptions, the importance sampling
algorithm is given by Fig.\ref{algo-imp-sam-cla}
(expressed in pseudo-code,
pidgin C language).

\begin{figure}[h]\begin{center}
\fbox{\parbox{5.5in}{
\begin{verse}
For all $(x)_H$ $\{W[(x)_H]=0;\}$\\
$W_{tot}=0;$\\
For samples $k=1,2, \ldots,N_{sam}\{$\\
\hspace{2em}$L=1;$\\
\hspace{2em}For nodes $i=1,2, \ldots,N_{nds}\{$\\
\hspace{4em}Generate $\sam{x_i}{k}$ from $Q(x_i|(\sam{x}{k})_{pa(i)});$\\
\hspace{4em}$//$Here, for LWS, $\sam{x_i}{k}==x_i$ when $i\in E$.\\
\hspace{4em}$//$$ pa(i)\subset Z_{1, i-1}$ so $(\sam{x}{k})_{pa(i)}$ known at this point.\\
\hspace{4em}if $i\in E$\{\\
\hspace{6em}if $\sam{x_i}{k}==x_i$\{\\
\hspace{8em}$L\;*=\;\frac{P(x_i|(\sam{x}{k})_{pa(i)})
}{Q(x_i|(\sam{x}{k})_{pa(i)})};$\\
\hspace{8em}$//$Here $\frac{P}{Q}= 1$ for RS and $\frac{P}{Q}=P$ for LWS.\\
\hspace{6em}\}else\{$//$LWS never enters here\\
\hspace{8em}go to next k;\\
\hspace{6em}$\}$\\
\hspace{4em}$\}$\\
\hspace{2em}$\}// i$ loop (nodes)\\
\hspace{2em}$W[(\sam{x}{k})_H]\;+=\;L;$ \\
\hspace{2em}$W_{tot}\;+=\;L;$\\
$\}// k$ loop (samples)\\
For all $(x)_H$ $\{P((x)_H|(x)_E)=\frac{W[(x)_H]}{W_{tot}};\}$\\
\end{verse}
}}
\caption{Algorithm for importance sampling of
 CB net on classical computer.}
 \label{algo-imp-sam-cla}
\end{center}
\end{figure}

\begin{claim}
For the algorithm of Fig.\ref{algo-imp-sam-cla},
$\frac{W[(x)_H]}{W_{tot}}\rarrow P((x)_H|(x)_E)$
as $N_{sam}\rarrow \infty$.
\end{claim}
\proof

Define the likelihood ratio function:

\beq
L_E(x') =
\prod_{i\in E}
\frac{P(x'_i|(x')_{pa(i)})
}{Q(x'_i|(x')_{pa(i)})}
\;
\eeq
for all $x'\in val(\rvx)$.
Clearly,

\beqa
Q(x')L_E(x') &=&
\prod_{i\in E^c}\{Q(x'_i|(x')_{pa(i)})\}
\prod_{i\in E}\{P(x'_i|(x')_{pa(i)})\}\\
&=& P(x')
\;.
\eeqa
For any function
$g:val(\rvx)\rarrow \RR$, as
$N_{sam}\rarrow \infty$,
the sample average
$\overline{g(\sam{x}{k})}$ tends to:

\beq
\overline{g(\sam{x}{k})} =
\frac{1}{N_{sam}}\sum_k g(\sam{x}{k})
\rarrow \sum_{x'}
Q(x')
\delta[(x)_E,(x')_E]
g(x')
\;.
\eeq
Therefore,

\beqa
\frac{W[(x)_H]}{W_{tot}}&=&
\frac{\frac{1}{N_{sam}}\sum_k L_E(\sam{x}{k})\delta[(x)_{H}, (\sam{x}{k})_{H}]}
{\frac{1}{N_{sam}}\sum_k L_E(\sam{x}{k})}\\
&\rarrow&
\frac{\sum_{x'}P(x') \delta[(x)_{E\cup H}, (x')_{E\cup H}]}
{\sum_{x'}P(x') \delta[(x)_E, (x')_E]}\\
&\rarrow&
\frac{P((x)_{E\cup H})}{P((x)_E)}
\;.
\eeqa
\qed

\section{Appendix: Markov Chain Monte Carlo\\for CB Nets
on a Classical Computer}

In this Appendix, we review
two examples (Gibbs and Metropolis-Hastings)
of
Markov Chain Monte Carlo (MCMC)
sampling algorithms
for CB nets
on a classical computer.

A {\bf Markov chain} is a CB net of the form

\beq
\rvx^0\rarrow\rvx^1\rarrow
\rvx^2\rarrow \ldots\rarrow \rvx^T
\;,
\eeq
where
$val(\rvx^t)$ is independent of
$t\in Z_{0,T}$.

It's clear from its
graph that a Markov chain satisfies

\beq
P(x^{t+1}| x^{t}, x^{t-1}, \ldots x^0) =
P(x^{t+1}| x^{t})
\;,
\eeq
i.e, the probability that $\rvx^{t+1}=x^{t+1}$
at time $t+1$ is independent of
what happened at all previous times except
at the immediate past $t$.
The $N_{\rvx^t}\times N_{\rvx^t}$ matrix
with entries $P(\rvx^{t+1}=x| \rvx^{t}=x')$ is called
the {\bf transition matrix} of the Markov chain;
we will represent it by $\tran$.
We will assume that
$\tran$ is independent of $t$
(this property of $\tran$ is called
time invariance or time homogeneity).

Let $\pi:val(\rvx^t)\rarrow \RR$ be a probability vector (
$\pi(x)\geq 0$ for all $x\in val(\rvx^t)$
and  $\sum_x \pi(x)=1$).

We say $\pi$
 is a
{\bf stationary distribution} of $\tran$
if

\beq
\tran\pi = \pi
\;,
\eeq
i.e.,
$\pi$ is an eigenvector of $\tran$
with unit eigenvalue.

We say $\pi$ is a
{\bf detailed balance} of $\tran$ if
$\tran(x'|x)\pi(x)$
is invariant under the exchange of $x$
and $x'$; that is,

\beq
\tran(x'|x)\pi(x) = \tran(x|x')\pi(x')
\;,
\label{eq-det-bal}
\eeq
for all $x,x'\in val(\rvx^t)$.
Detailed
balance is
tantamount to equilibrium  since
$\tran(x'|x)\pi(x)$ is the probability
flux being transmitted from state
$\rvx^t=x$ to state $\rvx^{t+1}=x'$
after a long time,
and $\tran(x|x')\pi(x')$ is that being
transmitted in the opposite direction,
and these two are equal. Hence,
it is not surprising that
if $\pi$ is a detailed balance of $\tran$,
then $\pi$ is a stationary distribution of $\tran$.
Indeed, summing over $x$
both sides of Eq.(\ref{eq-det-bal}) proves this.

A {\bf Markov Chain Monte Carlo (MCMC)
sampling algorithm} is a method whereby,
given a Markov chain
$\rvx^0\rarrow\rvx^1\rarrow\rvx^2\rarrow\ldots$, we
find a set of points in $val(\rvx^t)$
that is distributed according to the
stationary distribution $\pi$ of the Markov chain.
$\pi$ is taken to be the full probability
distribution of a CB net.
Next we
discuss two examples of MCMC sampling
algorithms:
the Gibbs and the Metropolis-Hastings
sampling algorithms. Actually,
the Gibbs algorithm is a
special case of the
Metropolis-Hastings one,
but I think it is
pedagogically
beneficial to
discuss the Gibbs
algorithm first, separately.

\subsection{Appendix:
Gibbs Sampling \\of CB Nets
on a Classical Computer}\label{app-gibbs-sam}

In this Appendix, we review
the Gibbs sampling algorithm
for CB nets
on a classical computer.

Consider a Markov chain
$\rvx^0\rarrow\rvx^1\rarrow\rvx^2\ldots \rarrow\rvx^T$.
Let
$\rvx^t = (\rvx_1^t, \rvx_2^t,\ldots,\rvx^t_{N_{nds}})$
for each time $t$
represent a separate copy of a CB net
with nodes $\rvx_1^t, \rvx_2^t,\ldots,\rvx^t_{N_{nds}}$,
and probability matrices
$P(x_i^t| (x^t)_{pa(i)})$.
(The nodes of the CB net
$\rvx^t$ are
 not necessarily in topological order.)
Assume that $E$ (evidence set)
and $H$ (hypotheses set)
are disjoint subsets of $Z_{1,N_{nds}}$,
with $Z_{1,N_{nds}}-E\cup H$ not
necessarily empty.
Let $X^c = Z_{1,N_{nds}}-X$ for any
$X\subset Z_{1,N_{nds}}$.
Assume that we are given
the prior evidence $(x)_E$.
All probabilities in this
section about the Gibbs algorithm
will be conditioned implicitly on
$(\rvx^t)_E = (x)_E$ for all $t$. The
Gibbs algorithm is designed to respect this constraint,
by never changing the value of $(\rvx^t)_E$
after it is initially set.
Assume that we are given the last time $T$ of the Markov chain,
and the burn time $t_{burn}$ ($0<<t_{burn}<<T$).

Under these assumptions, the Gibbs sampling
algorithm is given by Fig.\ref{algo-gibbs-sam-cla}
(expressed in pseudo-code,
pidgin C language).

\begin{figure}[h]\begin{center}
\fbox{\parbox{5in}{
\begin{verse}
For all $(x)_H$ $\{W[(x)_H]=0;\}$\\
$W_{tot}=0;$\\
Initialize $\rvx^0$ to some value $x^0$, subject to $(x^0)_E=(x)_E;$\\
For times $t=0,1,2, \ldots,T-1\{$\\
\hspace{2em}Draw $i$ uniformly from $Z_{1,N_{nds}}$;\\
\hspace{2em}if $i\in E$\{\\
\hspace{4em}$x_i^{t+1}=x_i$;\\
\hspace{2em}\}else\{\\
\hspace{4em}Generate $x_i^{t+1}\sim P(x_i^{t+1}|(x^t)_{MB(i)});$\\
\hspace{2em}$\}$\\
\hspace{2em}$(x^{t+1})_\noti = (x^t)_\noti;$\\
\hspace{2em}if $t> t_{burn}$\{$//$ $0<<t_{burn}<< T$\\
\hspace{4em}$W[(x^{t+1})_H]++;$ \\
\hspace{4em}$W_{tot}++;$\\
\hspace{2em}$\}$\\
$\}// t$ loop (times)\\
For all $(x)_H$ $\{P((x)_H|(x)_E)=\frac{W[(x)_H]}{W_{tot}};\}$\\
\end{verse}
}}
\caption{Algorithm for Gibbs sampling of
 CB net on classical computer. Nodes of CB net $\rvx^t$
 are visited at random.}
 \label{algo-gibbs-sam-cla}
 \end{center}
\end{figure}

\begin{claim} For the algorithm of Fig.\ref{algo-gibbs-sam-cla},
$P(\rvx^t=x^t)$ is a stationary distribution
of $P(\rvx^{t+1}=x^{t+1}|\rvx^t=x^t)$. In other words,

\beq
\sum_{x^t\in val(\rvx^t)}P(x^{t+1}|x^t)P(x^{t})=
P(x^{t+1})
\;
\label{eq-stat-dist}
\eeq
for all $x^{t+1}\in val(\rvx^t)$.
\end{claim}
\proof

One begins by conditioning
the transition matrix on the node index $i$:

\beq
P(x^{t+1}|x^t) =
\frac{1}{N_{nds}}
\sum_i P(x^{t+1}|x^t, i)
\;.
\eeq
Rather than proving Eq.(\ref{eq-stat-dist}),
we will prove the stronger statement

\beq
\sum_{x^t}P(x^{t+1}|x^t, i)P(x^{t})=
P(x^{t+1})
\;
\label{eq-stat-dist-i}
\eeq
for all $i\in Z_{1, N_{nds}}$.
If $P(x^{t+1})$
is a stationary distribution of
$P(x^{t+1}|x^t, i)=\tran(i)$
for any $i$, then
it is a stationary distribution
of any product
$\tran(i_1)\tran(i_2)\cdots\tran(i_n)$,
for any sequence $i_1,i_2,\ldots, i_n$
of $i$'s.

Studying the algorithm of Fig.\ref{algo-gibbs-sam-cla}
carefully,
we conclude that

\beq
P(x^{t+1}|x^t, i)
=
\left[\Theta(i\in E^c)P(x_i^{t+1}| (x^{t})_{MB(i)}
)
+\Theta(i\in E)\delta_{x_i^t}^{x_i^{t+1}}
\right]\delta_{(x^t)_\noti}^{(x^{t+1})_\noti}
\;.
\label{eq-gibbs-p-trans-i}
\eeq
For $i\in E$, $P(x^{t+1}|x^{t},i) = \delta_{x^t}^{x^{t+1}}$,
so Eq.(\ref{eq-stat-dist-i}) is clearly satisfied.
For $i\in E^c$,

\beqa
\sum_{x^t}P(x^{t+1}|x^t,i)P(x^t)
&=&
\sum_{x_i^t, (x^t)_\noti}
P(x_i^{t+1}| (x^t)_\noti)
\delta_{(x^t)_\noti}^{(x^{t+1})_\noti}
P(x^t)\\
&=& \sum_{x_i^t}
P(x_i^{t+1}|(x^{t+1})_\noti)
P((x^{t+1})_\noti, x^t_i)\\
&=&
P(x_i^{t+1}|(x^{t+1})_\noti)
P((x^{t+1})_\noti)\\
&=&
P(x^{t+1})
\;.
\eeqa
\qed

Rather than choosing nodes
of the graph $\rvx^t$ at random,
one can
sweep
through all of them,
in a fixed deterministic order,
repeating this all-nodes-sweep $N_{gra}$ times.
Hence, we can replace
the algorithm of Fig.\ref{algo-gibbs-sam-cla}
by
the one of Fig.\ref{algo-gibbs-sam-cla-det}.

\begin{figure}[h]\begin{center}
\fbox{\parbox{5.5in}{
\begin{verse}
For all $(x)_H$ $\{W[(x)_H]=0;\}$\\
$W_{tot}=0;$\\
Initialize $\rvx^0$ to some value $x^0$, subject to $(x^0)_E=(x)_E;$\\
$t=0;$\\
For graphs $g=1,2,3, \ldots,N_{gra}\{$\\
\hspace{2em}For nodes $i=1,2,\ldots,N_{nds} \{$;\\
\hspace{4em}if $i\in E$\{\\
\hspace{6em}$x_i^{t+1}=x_i$;\\
\hspace{4em}\}else\{\\
\hspace{6em}Generate $x_i^{t+1}\sim P(x_i^{t+1}|(x^t)_{MB(i)});$\\
\hspace{4em}$\}$\\
\hspace{4em}$(x^{t+1})_\noti = (x^t)_\noti;$\\
\hspace{4em}$t++;$\\
\hspace{2em}$\}// i$ loop (nodes)\\
\hspace{2em}if $t> t_{burn}$\{$//$ $0<<t_{burn}<< N_{nds}N_{gra}$\\
\hspace{4em}$W[(x^{t})_H]++;$ \\
\hspace{4em}$W_{tot}++;$\\
\hspace{2em}$\}$\\
$\}// g$ loop (graphs)\\
For all $(x)_H$ $\{P((x)_H|(x)_E)=\frac{W[(x)_H]}{W_{tot}};\}$\\
\end{verse}
}}
\caption{Algorithm for Gibbs sampling of
 CB net on classical computer.
 Nodes of CB net $\rvx^t$
 are visited in a fixed deterministic order. }
 \label{algo-gibbs-sam-cla-det}
\end{center}
\end{figure}

The sample of points
in $val(\rvx^t)$ generated by this ``nodes
in fixed order"
algorithm isn't time invariant
for $t$ differences $\Delta t =1$,
but is time invariant for
$\Delta t =N_{nds}$.

\subsection{Appendix:
Metropolis-Hastings Sampling \\of CB Nets
on a Classical Computer}\label{app-met-has-sam}

In this Appendix, we review
the Metropolis-Hastings sampling algorithm
for CB nets
on a classical computer.

Consider a Markov chain
$\rvx^0\rarrow\rvx_1\rarrow\rvx^2\ldots \rarrow\rvx^T$.
Let
$\rvx^t = (\rvx_1^t, \rvx_2^t,\ldots,\rvx^t_{N_{nds}})$
for each time $t$
represent a separate copy of a CB net
with nodes $\rvx_1^t, \rvx_2^t,\ldots,\rvx^t_{N_{nds}}$.
and probability matrices
$P(x_i^t| (x^t)_{pa(i)})$.
(The nodes of the CB net
$\rvx^t$ are
 not necessarily in topological order.)
Assume that $E$ (evidence set)
and $H$ (hypotheses set)
are disjoint subsets of $Z_{1,N_{nds}}$,
with $Z_{1,N_{nds}}-E\cup H$ not
necessarily empty.
Let $X^c = Z_{1,N_{nds}}-X$ for any
$X\subset Z_{1,N_{nds}}$.
Assume that we are given
the prior evidence $(x)_E$.
All probabilities in this
section about the Metropolis-Hastings algorithm
will be conditioned implicitly on
$(\rvx^t)_E = (x)_E$ for all $t$. The Metropolis-Hastings
algorithm is designed to respect this constraint,
by never changing the value of $(\rvx^t)_E$
after it is initially set.
Assume that we are given
the last time $T$ of the Markov chain,
the burn time $t_{burn}$ ($0<<t_{burn}<<T$),
and
{\bf sampling
probability distributions}
$Q_i(y_i|x^t_i, (x^t)_{MB(i)})$
where $i\in Z_{1,N_{nds}}$,
$y_i\in val(\rvx^t_i)$,
$x^t\in val(\rvx^t)$.

Under these assumptions, the Metropolis-Hastings sampling
algorithm is given by Fig.\ref{algo-met-has-cla}
(expressed in pseudo-code,
pidgin C language).

\begin{figure}[h]\begin{center}
\fbox{\parbox{6in}{
\begin{verse}
For all $(x)_H$ $\{W[(x)_H]=0;\}$\\
$W_{tot}=0;$\\
Initialize $\rvx^0$ to some value $x^0$, subject to $(x^0)_E=(x)_E;$\\
For times $t=0,1,2, \ldots,T-1\{$\\
\hspace{2em}Draw $i$ uniformly from $Z_{1,N_{nds}}$;\\
\hspace{2em}if $i\in E$\{\\
\hspace{4em}$x_i^{t+1}=x_i$;\\
\hspace{2em}\}else\{\\
\hspace{4em}Generate $y_i\sim Q_i(y_i|x_i^t,(x^t)_{MB(i)});$\\
\hspace{4em}Draw $u_i$ uniformly from the interval $[0,1];$\\
\hspace{4em}$\alpha_i = \min\left\{
1, \frac{
Q_i(x^t_i|y_i,(x^t)_{MB(i)} )P(y_i|(x^t)_{MB(i)})
}{
Q_i(y_i|x^t_i,(x^t)_{MB(i)})P(x^t_i|(x^t)_{MB(i)})
}
\right\};$\\
\hspace{4em}if
$(u_i<\alpha_i)\{ x_i^{t+1}=y_i;\}$
else
$\{x_i^{t+1}=x_i^t;\}$\\
\hspace{4em}$//$if $Q_i=P$, then $\alpha_i=\min(1,1)=1$, and get Gibbs\\
\hspace{2em}$\}$\\
\hspace{2em}$(x^{t+1})_\noti = (x^t)_\noti;$\\
\hspace{2em}if $t> t_{burn}$\{$//$ $0<<t_{burn}<< T$\\
\hspace{4em}$W[(x^{t+1})_H]++;$ \\
\hspace{4em}$W_{tot}++;$\\
\hspace{2em}$\}$\\
$\}// t$ loop (times)\\
For all $(x)_H$ $\{P((x)_H|(x)_E)=\frac{W[(x)_H]}{W_{tot}};\}$\\
\end{verse}
}}
\caption{Algorithm for Metropolis-Hastings sampling of
 CB net on classical computer.}
 \label{algo-met-has-cla}
\end{center}
\end{figure}

\begin{claim}\label{cl-met-has}
For the algorithm of Fig.\ref{algo-met-has-cla},
$P(\rvx^t=x^t)$ is a stationary distribution
of $P(\rvx^{t+1}=x^{t+1}|\rvx^t=x^t)$. In other words,

\beq
\sum_{x^t\in val(\rvx^t)}P(x^{t+1}|x^t)P(x^{t})=
P(x^{t+1})
\;
\label{eq-stat-dist-v2}
\eeq
for all $x^{t+1}\in val(\rvx^t)$.
\end{claim}
\proof

One begins by conditioning
the transition matrix on the node index $i$:

\beq
P(x^{t+1}|x^t) =
\frac{1}{N_{nds}}
\sum_i P(x^{t+1}|x^t, i)
\;.
\eeq
Rather than proving Eq.(\ref{eq-stat-dist-v2}),
we will prove the stronger statement

\beq
\sum_{x^t}P(x^{t+1}|x^t, i)P(x^{t})=
P(x^{t+1})
\;
\label{eq-stat-dist-i-v2}
\eeq
for all $i\in Z_{1, N_{nds}}$.
If $P(x^{t+1})$
is a stationary distribution of
$P(x^{t+1}|x^t, i)=\tran(i)$
for any $i$, then
it is a stationary distribution
of any product
$\tran(i_1)\tran(i_2)\cdots\tran(i_n)$,
for any sequence $i_1,i_2,\ldots, i_n$
of $i$'s.

\begin{figure}[h]
    \begin{center}
    \epsfig{file=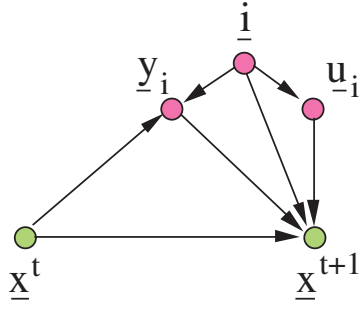, height=1.7in}
    \caption{CB net connecting the
    random variables used in the Metropolis-Hastings
    algorithm.}
    \label{fig-met-has-cbnet}
    \end{center}
\end{figure}

Studying the algorithm
of Fig.\ref{algo-met-has-cla} carefully,
we conclude the following.
For $i\in E$, $P(x^{t+1}|x^{t},i) = \delta_{x^t}^{x^{t+1}}$,
so Eq.(\ref{eq-stat-dist-i-v2}) is clearly satisfied.
For $i\in E^c$,

\beq
P(x^{t+1}|x^t, i) =
\sum_{y_i}\int_0^1 du_i
P(x^{t+1}|x^t, y_i, u_i,i)
P(y_i|x^t,i)
P(u_i|i)
\;.
\label{eq-met-has-net}
\eeq
Eq.(\ref{eq-met-has-net})
comes from
the CB net of Fig.\ref{fig-met-has-cbnet}.
The 3 probabilities
occurring on the right hand side
of Eq.(\ref{eq-met-has-net}) are
given by

\begin{subequations}
\beq
P(x^{t+1}|x^t, y_i, u_i,i)=
\delta^{(x^t)_\noti}_{(x^{t+1})_\noti}
\;\;
\delta_{x_i^{t+1}}^{y_i\Theta(u_i<\alpha_i)
+x_i^t\Theta(u_i>\alpha_i)}
\;,
\eeq

\beq
P(y_i|x^t,i)= Q_i(y_i|x^t_i,(x^t)_{MB(i)})
\;,
\eeq
and

\beq
P(u_i|i) = 1
\;.
\eeq
\end{subequations}
($u_i$ is a continuous random
variable.)

One can ``sum over" the node $u_i$
 of
Fig.\ref{fig-met-has-cbnet}:

\beq
\int_0^1 du_i P(x^{t+1}|x^t, y_i, u_i,i)=
\delta^{(x^t)_\noti}_{(x^{t+1})_\noti}
\left[
\delta_{x_i^{t+1}}^{y_i}\alpha_i
+
\delta_{x_i^{t+1}}^{x_i^t}(1-\alpha_i)
\right]
\;.
\eeq
Define

\beq
\qbar_i(y_i|x^t)
\equiv \alpha_i Q_i(y_i|x^t_i,(x^t)_{MB(i)})
\;.
\eeq
One can also sum over the node $y_i$ of
Fig.\ref{fig-met-has-cbnet}:

\beqa
P(x^{t+1}|x^t, i) &=&
\sum_{y_i}
\delta^{(x^t)_\noti}_{(x^{t+1})_\noti}
\left[
\delta_{x_i^{t+1}}^{y_i}\alpha_i
+
\delta_{x_i^{t+1}}^{x_i^t}(1-\alpha_i)
\right]Q_i(y_i|x^t_i,(x^t)_{MB(i)})\\
&=&
\delta^{(x^t)_\noti}_{(x^{t+1})_\noti}
\left\{
\qbar_i(x_i^{t+1}|x^t)
+
\delta^{x_i^t}_{x_i^{t+1}}
[1-\sum_{y_i}\qbar_i(y_i|x^t)]
\right\}
\;.
\label{eq-met-has-p-trans-i}
\eeqa

Note that
\beqa
P(x^t) &=&
P(x_i^t|(x^t)_\noti)P((x^t)_\noti)\\
&=&
P(x_i^t|(x^t)_{MB(i)})P((x^t)_\noti)
\;.
\eeqa
Hence,

\beqa
\scriptstyle
\qbar(y_i|x^t)P(x^t)
&=&
\scriptstyle
\min\left\{
1, \frac{
Q_i(x^t_i|y_i,(x^t)_{MB(i)})P(y_i|(x^t)_{MB(i)})
}{
Q_i(y_i|x^t_i,(x^t)_{MB(i)})P(x^t_i|(x^t)_{MB(i)})
}
\right\}
Q_i(y_i|x^t_i,(x^t)_{MB(i)})
P(x^t)\nonumber\\
&=&
\scriptstyle
\min\left\{
\begin{array}{l}\scriptstyle
Q_i(y_i|x^t_i,(x^t)_{MB(i)})P(x^t_i|(x^t)_{MB(i)}),
\\
\scriptstyle
Q_i(x_i^t|y_i,(x^t)_{MB(i)})P(y_i|(x^t)_{MB(i)})
\end{array}
\right\}P((x^t)_\noti)
\;.
\label{eq-qbar-det-bal}
\eeqa
From Eq.(\ref{eq-qbar-det-bal}), we
see that  $\qbar(y_i|x^t)P(x^t)$
is invariant under the exchange of
$y_i$ and $x^t_i$.
In other words,
$P(x^t)$
is a detailed balance
of $\qbar(y_i|x^t)$
under the exchange of
$y_i$ and $x^t_i$.
However, $\sum_{y_i}\qbar(y_i|x^t)\neq 1$,
so $\qbar(y_i|x^t)$ is not a
probability distribution in $y_i$.

Now we have

\beqa
\sum_{x^t}P(x^{t+1}|x^t,i)P(x^t)&=&
 \sum_{x^t}
\delta^{(x^t)_\noti}_{(x^{t+1})_\noti}
\left\{
\begin{array}{c}
\qbar_i(x_i^{t+1}|x^t)\\
+\\
\delta^{x_i^t}_{x_i^{t+1}}
[1-\sum_{y_i}\qbar_i(y_i|x^t)]
\end{array}
\right\}
P(x^t)\nonumber\\
&=& T_1 + T_2 + T_3
\;,
\eeqa
where

\beqa
T_1&=&\sum_{x^t}
\delta^{(x^t)_\noti}_{(x^{t+1})_\noti}
\qbar_i(x_i^{t+1}|x^t)
P(x^t)\\
&=&\sum_{x_i^t}
\qbar_i(x_i^{t+1}|x_i^t,(x^{t+1})_\noti)
P(x_i^t,(x^{t+1})_\noti)
\label{eq-pre-det-bal}\\
&=&\sum_{x_i^t}
\qbar_i(x_i^t|x_i^{t+1},(x^{t+1})_\noti)
P(x_i^{t+1},(x^{t+1})_\noti)
\label{eq-post-det-bal}\\
&=&\sum_{x_i^t}
\qbar_i(x_i^t|x^{t+1})
P(x^{t+1})
\;,
\eeqa
(To go from Eq.(\ref{eq-pre-det-bal})
to Eq.(\ref{eq-post-det-bal}),
we used the fact that
$P(x^t)$
is a detailed balance
of $\qbar(y_i|x^t)$.)

\beq
T_2=\sum_{x^t}
\delta^{x^t}_{x^{t+1}}
P(x^t)=P(x^{t+1})
\;,
\eeq
and

\beqa
T_3&=& -\sum_{x^t}
\delta^{x^t}_{x^{t+1}}
\sum_{y_i}\qbar_i(y_i|x^t)
P(x^t)\\
&=&-\sum_{y_i}\qbar_i(y_i|x^{t+1})
P(x^{t+1})\\
&=& -T_1
\;.
\eeqa
\qed

As we pointed out for the Gibbs
algorithm in Appendix \ref{app-gibbs-sam},
rather than choosing nodes
of the graph $\rvx^t$ at random,
one can
sweep through all the nodes of the graph
in a fixed order,
repeating this sweep
$N_{gra}$ times.
Hence, we can replace
the algorithm of Fig.\ref{algo-met-has-cla} by
an alternative one.
We leave the details to the reader.

There are several important special cases
of the algorithm of Fig.\ref{algo-met-has-cla}:
\begin{enumerate}
\item
When $Q_i(y_i|x_i^t, (x^t)_{MB(i)}) = P(y_i|(x^t)_{MB(i)})$,
we get $\alpha_i=\min(1,1)=1$,
yielding the Gibbs algorithm
discussed in Appendix \ref{app-gibbs-sam}.

\item When $N_{nds}=1$, there is no need for $i$ subscripts
in $x^t_i$ or $y_i$.
There is also no possibility of evidence.
Hence, the algorithm of Fig.\ref{algo-met-has-cla}
simplifies to the one in Fig.\ref{algo-met-has-cla-one-node}.

\begin{figure}[h]\begin{center}
\fbox{\parbox{5in}{
\begin{verse}
For all $(x)_H$ $\{W[(x)_H]=0;\}$\\
$W_{tot}=0;$\\
Initialize $\rvx^0$ to some value $x^0;$\\
For times $t=0,1,2, \ldots,T-1\{$\\
\hspace{2em}Generate $y\sim Q(y|x^t);$\\
\hspace{2em}Draw $u$ uniformly from the interval $[0,1];$\\
\hspace{2em}$\alpha = \min\left\{
1, \frac{
Q(x^t|y)P(y)
}{
Q(y|x^t)P(x^t)
}
\right\};$\\
\hspace{2em}if
$(u<\alpha)\{ x^{t+1}=y;\}$
else
$\{x^{t+1}=x^t;\}$\\
\hspace{2em}if $t> t_{burn}$\{$//$ $0<<t_{burn}<< T$\\
\hspace{4em}$W[(x^{t+1})_H]++;$ \\
\hspace{4em}$W_{tot}++;$\\
\hspace{2em}$\}$\\
$\}// t$ loop (times)\\
For all $(x)_H$ $\{P((x)_H)=\frac{W[(x)_H]}{W_{tot}};\}$\\
\end{verse}
}}
\caption{Algorithm for Metropolis-Hastings sampling of
 CB net on classical computer. Special case
 where $N_{nds}=1$.}
 \label{algo-met-has-cla-one-node}
\end{center}
\end{figure}

\item If $Q_i(x_i^t|y_i, (x^t)_{MB(i)})$
is invariant under the exchange of
$x_i^t$ and $y_i$, then

\beq
\alpha_i = \min \left\{ 1,
\frac{P(y_i|(x^t)_{MB(i)})
}{
P(x^t_i|(x^t)_{MB(i)})
}
\right\}
\;.
\eeq
In particular, if $N_{nds}=1$,

\beq
\alpha= \min \left\{ 1,
\frac{P(y)
}{
P(x^t)
}
\right\}
\;.
\eeq
This is called the
{\bf Metropolis sampling algorithm}.
It was invented prior to the
Metropolis-Hastings one.
\end{enumerate}

The algorithm of Fig.\ref{algo-met-has-cla}
and the proof of Claim \ref{cl-met-has}
are both fairly complicated.
One wonders, why do they work?
What logic motivated the invention
of the algorithm?
Here is an intuitive explanation of
what is going on.
Define

\beq
\fluxtox \equiv Q_i(x_i^t|y_i,(x^t)_{MB(i)})
P(y_i|(x^t)_{MB(i)})
\;.
\label{eq-flux-def}
\eeq
Define $\fluxtoy$ as the expression on
the right hand side
of Eq.(\ref{eq-flux-def}), but
$x_i^t$ and $y_i$ exchanged.
$\fluxtox$ is the probability
flux flowing
from state $y_i$ to state $x_i^t$, and
$\fluxtoy$ is the flux in the opposite
direction. Now $\alpha_i$, often called
the ``acceptance probability",
can be expressed as

\beq
\alpha_i = \min\left\{ 1, \frac{\fluxtox}{\fluxtoy}\right\}
\;.
\eeq
Note that this definition of $\alpha_i$
puts it in the interval $[0,1]$, as
required for a probability. Recall that
the algorithm defines:

\beq
x^{t+1}_i = y_i\Theta(u_i<\alpha_i)
+ x_i^t \Theta(u_i>\alpha_i)
\;.
\eeq
Thus,

\begin{enumerate}
\item If $\fluxtox<<\fluxtoy$, then
$\alpha_i = \fluxtox/\fluxtoy<< 1$. In this
case, we
\begin{enumerate}
\item \label{it-infreq}
Set $x_i^{t+1}=y_i$ (i.e., accept the new value),
doing this
infrequently, $\alpha_i$ of
the time.
\item
Set
 $x_i^{t+1}=x_i^t$ (i.e., keep the old value),
 doing this frequently,
 $1-\alpha_i$ of the time.
 \end{enumerate}
\item If $\fluxtox>\fluxtoy$, then
$\alpha_i = 1$. In this
case, we always set
$x_i^{t+1}=y_i$ (i.e., accept the new value).
\end{enumerate}
Most of the time (except
in case \ref{it-infreq}), we are trying to ``buck (counteract) the
trend" that state $x_i^t$ is either gaining
or loosing weight.
We don't buck the trend always,
because we want to allow a small probability
of escaping local minima.

\section{Appendix: Quantum Circuits for\\
Two Examples of Ref.\cite{Tuc00}}
\label{app-quan-ckts}

\begin{figure}[t]
    \begin{center}
    \epsfig{file=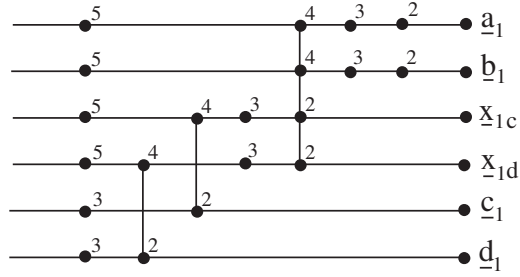, height=1.5in}
    \caption{Quantum circuit for two body scattering QB net of
    Ref.\cite{Tuc00}}
    \label{fig-scattering-ckt}
    \end{center}
\end{figure}

\begin{figure}[h]
    \begin{center}
    \epsfig{file=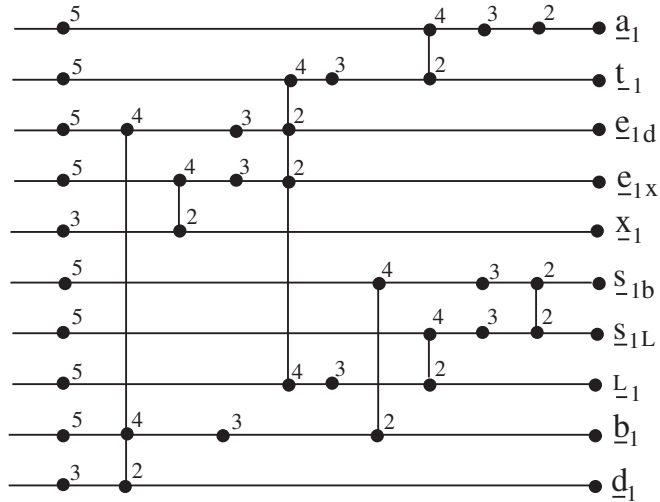, height=2.7in}
    \caption{Quantum circuit for Asia QB net of
    Ref.\cite{Tuc00}}
    \label{fig-asia-ckt}
    \end{center}
\end{figure}

Figs.\ref{fig-scattering-ckt}
and \ref{fig-asia-ckt}
are quantum circuits for the
two-body scattering and Asia nets,
respectively, that are discussed in
Ref.\cite{Tuc00}.

\end{appendix}


\begin{thebibliography}{99}


\bibitem{inv-trans}

http://en.wikipedia.org/wiki/Inverse\_transform\_sampling


\bibitem{arm}

http://en.wikipedia.org/wiki/Rejection\_sampling


\bibitem{Jordan}
M. Jordan (Editor) {\it Learning
in Graphical Models} (1999, The MIT Press)

\bibitem{Tuc00}
R.R. Tucci,
``Quantum Computer as a Probabilistic
Inference Engine",
arXiv:quant-ph/0004028 v2 .

\bibitem{Tuc-QMR}
R.R. Tucci,
``Use of a Quantum Computer and the Quick Medical Reference
To Give an Approximate Diagnosis",
arXiv:0806.3949

\bibitem{grover}
Lov K. Grover,
``Rapid sampling through quantum computing",
quant-ph/9912001

\bibitem{woc1}
Pawel Wocjan, Anura Abeyesinghe,
``Speed-up via Quantum Sampling",
arXiv:0804.4259

\bibitem{woc2}
Pawel Wocjan, Chen-Fu Chiang, Anura Abeyesinghe, Daniel Nagaj,
``Quantum Speed-up for Approximating Partition Functions",
arXiv:0811.0596

\bibitem{fox}Charles Fox,
``An Entangled Bayesian Gestalt: Mean-field, Monte-Carlo and Quantum Inference in Hierarchical Perception",
 DPhil thesis, Oxford, 2008

 \bibitem{Paulinesia}
R.R.Tucci,
``QC Paulinesia",
quant-ph/0407215



\bibitem{Golub}
G.H. Golub and C.F. Van Loan,
{\it Matrix Computations, Third Edition}
(John Hopkins Univ. Press, 1996).

\bibitem{Tuc99}
R.R. Tucci,
``A Rudimentary Quantum
Compiler(2cnd Ed.)",
arXiv:quant-ph/9902062



\bibitem{chain}
Ville Bergholm, Juha J. Vartiainen, Mikko Mottonen, Martti M. Salomaa.
``Quantum circuits with uniformly controlled one-qubit gates",
arXiv:quant-ph/0410066

\end{thebibliography}
\end{document}